\documentclass [12pt, twoside] {uwthesis}
 
\usepackage{graphicx,amssymb,amsmath,natbib}
\usepackage[figuresright]{rotating}

\newcommand{\fsize}{0.9\textwidth}
\newcommand{\cs}{$\chi^2$}

\begin{document}

%

\prelimpages
 
%
%
\Title{Detecting New Planets in Transiting Systems}
\Author{Jason Steffen}
\Year{2006}
\Program{Physics}

{\Degreetext{A dissertation
  submitted in partial fulfillment of\\
  the requirements for the degree of}
 \def\thefootnote{\fnsymbol{footnote}}
 \let\footnoterule\relax
 \titlepage
 }
\setcounter{footnote}{0}
 
%
%

\Chair{Eric Agol}{Professor}{Physics}
\Signature{Eric Agol}
\Signature{Paul Boynton}
\Signature{Craig Hogan}
\signaturepage

%




 

\doctoralquoteslip


%
%

\setcounter{page}{-1}
\abstract{
I present an initial investigation into a new planet detection technique that uses the transit timing of a known, transiting planet.  The transits of a solitary planet orbiting a star occur at equally spaced intervals in time.  If a second planet is present, dynamical interactions within the system will cause the time interval between transits to vary.  These transit time variations can be used to infer the orbital elements of the unseen, perturbing planet.  I show analytic expressions for the amplitude of the transit time variations in several limiting cases.  Under certain conditions the transit time variations can be comparable to the period of the transiting planet.  I also present the application of this planet detection technique to existing transit observations of the TrES-1 and HD209458 systems.  While no convincing evidence for a second planet in either system was found from those data, I constrain the mass that a perturbing planet could have as a function of the semi-major axis ratio of the two planets and the eccentricity of the perturbing planet. Near low-order, mean-motion resonances (within about 1\% fractional deviation), I find that a secondary planet must generally have a mass comparable to or less than the mass of the Earth--showing that these data are the first to have sensitivity to sub Earth-mass planets orbiting main sequence stars.  These results show that TTV will be an important tool in the detection and characterization of extrasolar planetary systems.
}
 
%
%
\tableofcontents

\vskip0.1in
\noindent
Bibliography
\hskip2.0pt
\hskip1.5pt .
\hskip1.5pt .
\hskip1.5pt .
\hskip1.5pt .
\hskip1.5pt .
\hskip1.5pt .
\hskip1.5pt .
\hskip1.5pt .
\hskip1.5pt .
\hskip1.5pt .
\hskip1.5pt .
\hskip1.5pt .
\hskip1.5pt .
\hskip1.5pt .
\hskip1.5pt .
\hskip1.5pt .
\hskip1.5pt .
\hskip1.5pt .
\hskip1.5pt .
\hskip1.5pt .
\hskip1.5pt .
\hskip1.5pt .
\hskip1.5pt .
\hskip1.5pt .
\hskip1.5pt .
\hskip1.5pt .
\hskip1.5pt .
\hskip1.5pt .
\hskip1.5pt .
\hskip1.5pt .
\hskip1.5pt .
\hskip1.5pt .
\hskip1.5pt .
\hskip1.5pt .
\hskip12pt
97


\listoffigures
\listoftables  
 
%
%

 
%
%
\acknowledgments{
  {\narrower\noindent
Many people deserve acknowledgement for the contributions that they made in my life.  Most will be left out by this list.  However, I would be ungrateful if I did not hit some highlights.  In chronological order, I would like to thank:

\noindent $\bullet$ My parents for supporting me in developing any talent that I may have had

\noindent $\bullet$ My siblings for their example throughout my life

\noindent $\bullet$ Davis High School for the math and the drumline

\noindent $\bullet$ Markus Cleverley and the missionaries for teaching me how to improve myself

\noindent $\bullet$ Weber State University Physics Department for being a diamond in the rough

\noindent $\bullet$ Faith Kimball Steffen for encouraging me to reach, for joining in the ride, and for making the future possible

\noindent $\bullet$ Wick Haxton and Martin Savage for their invaluable advice, Summer 1999

\noindent $\bullet$ Michael Kesden and the 1999 REU interns for showing me the possibilities

\noindent $\bullet$ Julianne Dalcanton for starting my research and for directing me to Eric

\noindent $\bullet$ Paul Boynton for his patience, guidance, and encouragement

\noindent $\bullet$ Michael Moore (not the film maker) and the Gravity Group for showing me the many wierd ways to view different problems

\noindent $\bullet$ My advisor, Eric Agol, for giving me an unbeatable thesis topic and liberty in its development

\noindent $\bullet$ James Steffen for his ability to make me smile at 3:30am.

\noindent $\bullet$ My Father for his direction.  It is He, above all, who has made me who I am

  \par}
}

%
%
\dedication{\begin{center}to my family, immediate and extended\end{center}}

%

%
%

\textpages
 
\chapter{Introduction}

Since the first discovery, a decade ago, of a planet orbiting a distant main-sequence star~\citep{mayor95}, a new field of astronomy has emerged with the potential to address
fundamental questions about our own solar system since we can
now compare it with other planetary systems.  
Several search techniques have been employed to identify additional planets (``exoplanets'') in orbit around distant stars.  The primary mode 
for discovery of exoplanets has been the measurement of the stellar radial velocities
via the Doppler effect upon the spectral lines of the host star.  Currently the small reflex motion of the star
due to the orbiting planet can only be detected for $m_{planet}\gtrsim 
10 m_\oplus$ \citep{but04,mca04,san04}.  
Other planned or existing techniques include astrometry \citep{for03,glind03,pravdo04,gouda05,lattanzi05}, planetary microlensing~\citep{albrow00,tsapras03}, and direct imaging \citep{per00, cha03,for03,bor03a,gou04}.  Recently, a large number of
planetary transit searches are being carried out which
are starting to yield an handful of giant planets 
\citep{cha00,kon03,pon04,kon04,bou04, alo04}, and many more planned 
searches should reap a large harvest of transiting planets in the
near future \citep{hor03}.  Despite these successes, the discovery of 
``terrestrial'' extrasolar planets, similar in size and mass to the Earth, awaits further developments in all of these planet detection techniques.

In this dissertation I present much of the founding research of a recent addition to the repertoire of planet search techniques which consists of looking for additional planets in a known, transiting system by analyzing the variation in the time between planetary transits.  These transit timing variations (TTV), which are caused by the gravitational perturbation of a secondary planet, can be used to constrain the orbital elements of the unseen, perturbing planet---even if its mass is comparable to the mass of the Earth~\citep{mir02,agol05,holm05}.  For the near term, this new technique can probe for planets around main-sequence stars that are too small to detect by any other means.  This sensitivity is particularly manifest near mean-motion resonances, which recent works by \cite{thommes}, \cite{papal}, and \cite{zhou} suggest might be very common.

The application of TTV depends upon the discovery and monitoring of transiting planetary systems.  The first successful detection of a planetary transit~\citep{cha00,henry00} was for a planet that had originally been identified from the reflex motion of the primary star, HD209458b~\citep{mazeh00}.  Since the mass of the planet is degenerate with orbital inclination, the planetary status of
the companion was confirmed once the transits were seen (since planetary transits imply that the system is seen edge-on).  {\it HST} observations of the HD209458 yielded precision measurements of the transit
lightcurve \citep{bro01} and made this the surest planetary
candidate around a main sequence star (other than our own).

The first extrasolar planet to be discovered primarily from transit data was the OGLE-TR-56b system reported by \cite{kon03}.  Existing and future searches for planetary transits, such as the COROT~\citep{bourde03}, XO \citep{mcc05}, and Kepler~\citep{borucki} missions, are expected to identify many, possibly hundreds, of transiting systems.  Each of the systems that will be discovered is a candidate for an analysis of the timing of the planetary transits.  These analyses may yield additional planetary discoveries or may provide important constraints on the presence of additional bodies in each system.  One such system is the TrES-1 planetary system, reported by \cite{alo04}, which was also discovered via planetary transits.  In a recent paper by David Charbonneau and collaborators~\citep{char05} the detection of thermal emission from the surface of TrES-1b was announced.  This paper includes the timing of 12 planetary transits.  In chapter 
5, I present the results from the first published TTV analysis of transit data that was designed to identify and constrain potential companion planets in such a system.  The analysis of these data proved to be the first probe for sub earth-mass planets in orbit about a distant, main-sequence star.

Beyond planet detection, TTV has other applications that are important for extrasolar planet research.  For example, while the ratio
of the planetary radius to the stellar radius of a transiting system can be measured with
extreme precision \citep{man02}, the absolute radii remain
uncertain due to a degeneracy that exists between the radius and the mass of the host star
\citep{sea03a}---an increase in the mass and radius of the star can
yield an identical lightcurve and period.  This mass-radius degeneracy may be broken via TTV provided there is an additional planet in the system.
About 10 per cent of the stars with known planetary companions have more than 
one planet, while possibly as much as 50 per cent of 
them show a trend in radial velocity indicative of
additional planets \citep{fis01}.
If one or both of the planets is transiting, dynamical interactions
between the planets will alter the timing of the transits
{\citep{dob96,cha00,mir02}.}  A measurement of these timing
variations, combined with radial velocity data, can break the
mass-radius degeneracy.

Other applications of TTV have significant ramifications for various theories of the formation and evolution of multiple planet systems.  One example is that TTV may be used to identify the relative inclinations of planetary systems.  A sample of relative inclinations of planetary systems would constrain the mechanisms by which the eccentricities of planets can grow~\citep{rafikov03}.  A second example is that TTV is well suited to detect the presence of small, rocky planets that may be trapped in low-order, mean-motion resonances.  The presence of these close-in, resonant, terrestrial planets favors a sequential-accretion model of planet formation over a gravitational instability model~\citep{zhou}.

Given the multiple motivations of detecting terrestrial planets, breaking
the mass-radius degeneracy, and constraining planetary system formation and evolution theories, the results of this work, along with those of planned developments, should prove to be an important tool for this new field of extrasolar planetary science.  In this dissertation I present analytic and numerical results 
for transit timing variations due to the presence of a second planet in chapter 
2.  That chapter shows the transit timing variations that are expected in several limiting cases such as non-interacting planets, an eccentric exterior perturbing planet with a large period relative to the transiting planet, the general transit timing differences for two planets
with circular, co-planar orbits, the case of exact 
mean-motion resonance, and the case of two eccentric planets.  I also present the results from numerical simulations of several known
multi-planet systems (though they are not transiting).

In chapter 
3, I discuss in more detail some of the applications of the TTV technique.  I present the results of a preliminary study to characterize the efficiency with which the TTV technique can be applied to discover secondary planets in transiting systems in chapter 
4.  Chapters 
5 and 6 contain results of TTV analyses of the transit times of the known, transiting systems TrES-1 and HD209458.  Finally, I give some concluding remarks in chapter 
7.

\chapter{The TTV Signal\label{analytic}}

In this chapter I present various mechanisms that can cause variations in the period of a transiting planet.  In general, all of these mechanisms are present in a physical system.  However, a study of several limiting cases serves both to identify important relationships among the orbital elements in relevant systems and to identify the effects of each individual means of perturbation.  Much of this discussion can be found in \cite{agol05}.

I will begin by outlining the coordinate system that I use for these analytic treatements.  I then discuss, in turn, the effects of: noninteracting planets (\S \ref{noninteracting}), a perturbing planet on a wide, eccentric orbit (\S \ref{largeecc}), non-resonant planets with perturbatively small eccentricities (\S \ref{nonresonant}), resonant systems (\S \ref{resonant}), and orbits with arbitrary eccentricities (\S \ref{generalorbs}).  I note that this treatment addresses only systems where the orbits are coplanar.

Throughout the rest of this work I characterize the strength
of transit timing variations as follows.
For a series of transit times, $t_j$, I fit the times assuming
a constant period, $P$ and compute the standard deviation, 
$\sigma$, of the difference between the nominal and actual times.  
Mathematically,
\begin{equation}
\sigma=\left[\frac{1}{N}\sum_{j=0}^{N-1} (t_j-t_0+Pj)^2\right]^{1/2},
\end{equation}
where $P$ and $t_0$ are chosen to minimize $\sigma$.
If the variations are strictly
sinusoidal, then the amplitude of the timing deviation is
simply $\sqrt{2}$ times larger than $\sigma$.

\section{The Coordinate Systems\label{coors}}

For this study of three body systems the positions of the objects, with arbitrary origin, are given by the Cartesian coordinates $\textbf{R}_i,\ i=0,1,2$ where the labels correspond to the central body and the two planetary companions respectively.  According to Newton's law of gravity, the accelerations of the masses are given by
\begin{equation}
\ddot{\bf R}_i = \sum_{j\ne i} G m_j\frac{{\bf R}_j-{\bf R}_i}{\vert{\bf R}_j-{\bf R}_i\vert^3}.
\label{carteq}
\end{equation}

By multiplying the equation for each particle by that particle's mass and adding them together, one finds:
\begin{equation}
m_0 \ddot{\bf R}_0+m_1 \ddot{\bf R}_1+m_2 \ddot{\bf R}_2 = 0.
\end{equation}
which states that the center of mass of the system experiences no external forces. Since light travel time and
parallax effects are negligible (see \S \ref{lighttravel}), the timing of each transit is unaffected by the total velocity or position of the center of mass.  So, setting 
\begin{equation}
{\bf R}_{cm}\equiv \frac{\sum_{i=0}^2 m_i {\bf R}_i}{\sum_{i=0}^2 m_i} = 0
\end{equation}
reduces the differential equations
of motion to two, which I take to be that of the two planets, $R_1$
and $R_2$.

When I solve these equations of motion numerically I use this Cartesian coordinate system.  However, for an analytic investigation it is more convenient to write the problem in Jacobian coordinates, a coordinate system that is commonly used in perturbation theory
for many bodies \citep[see, e.g.][]{mur99, mal93a, mal93b}.  For the
three-body problem, the Jacobian coordinates amount to three new
coordinates which describe (a) the center of mass of the system; (b)
the relative position of inner planet and the star (the ``inner
binary''); and (c) the relative position of the outer planet and the
barycenter of the inner binary (the ``outer binary'').  To distinguish
from the Cartesian coordinates, I denote the Jacobian coordinates with a
lower case ${\bf r}_i$.  The Jacobian coordinates are
\begin{eqnarray}
{\bf r}_0 &=& {\bf R}_{cm} = 0,\cr
{\bf r}_1 &=& {\bf R}_1 - {\bf R}_0,\cr
{\bf r}_2 &=& {\bf R}_2 - \frac{1}{m_0+m_1}\left[m_0{\bf R}_0+m_1{\bf R}_1\right].
\end{eqnarray}
Using  $\mu_i = m_i/M \sim m_i/m_0$, where $M=\sum_{i=0}^2 m_i$,
the equations of motion (\ref{carteq}) may be rewritten in Jacobian coordinates,
\begin{eqnarray}\label{jaceq}
\ddot{{\bf r}}_1 &=& -\frac{Gm_0}{1-\mu_1}\frac{{\bf r}_1}{r_1^3} 
- GM\mu_2 \frac{{\bf r}_1 - {\bf r}_{21}}{\vert {\bf r}_1 -{\bf r}_{21}\vert^3}
- GM\mu_2 \frac{{\bf r}_{21}}{r_{21}^3 },\cr
\ddot{{\bf r}}_2 &=& -\frac{Gm_0}{1-\mu_2}\frac{{\bf r}_{21}}{r_{21}^3}
- GM\mu_1 \frac{{\bf r}_{21} - {\bf r}_1}{\vert{\bf r}_{21} -{\bf r}_1\vert^3},
\end{eqnarray}
where ${\bf r}_{21}=\mu_1{\bf r}_1+{\bf r}_2={\bf R}_2-{\bf R}_0$.

\section{An interior, noninteracting planet\label{noninteracting}}

\begin{figure}
\includegraphics[width=\fsize]{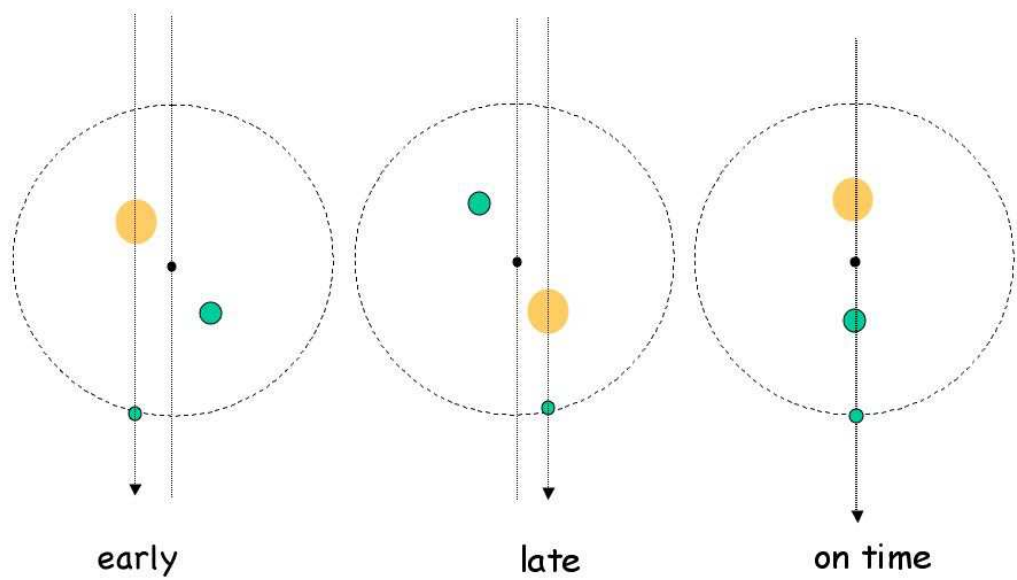}
\caption{Cartoon showing how an interior planet that does not directly interact with an exterior, transiting planet can change the timing of transits of the exterior planet.}
\label{fig1}
\end{figure}

The first, limiting case that I study assumes that the interaction between the planets is negligible.  At first glance it may seem that such an investigation would be fruitless and, indeed, if the interior planet is the one that transits there are no variations in the transit times due to a noninteracting secondary planet.  However, the motion of the inner binary causes the position of the star to oscillate with respect to the stationary center of mass of the system.  This oscillation will, in turn, cause the transits of the outer planet to vary as illustrated in figure \ref{fig1}.

To quantify these deviations, we take the limit as $\mu_1,\mu_2 \rightarrow 0$ in equation (\ref{jaceq}) which gives the equations of motion
\begin{eqnarray}\label{jackepler}
\ddot{{\bf r}}_1 &=& -Gm_0\frac{{\bf r}_1}{r_1^3},\cr
\ddot{{\bf r}}_2 &=& -Gm_0\frac{{\bf r}_2}{r_2^3}.
\end{eqnarray}
This approximation requires that the periapse of the
outer planet be much larger than the apoapse of the inner planet,
$(1-e_2)a_2\gg (1+e_1)a_1$ where $a_1, a_2$ are the semi-major axes of
the inner and outer binary and $e_1, e_2$ are the eccentricities.  The solution to these equations are standard Keplerian trajectories.

The simplest case to consider is that in which both the inner and
outer binary are on approximately circular orbits.  The transit occurs
when the outer planet is nearly aligned with the barycenter of the
inner binary and its motion during the transit is
essentially transverse to the line of sight.
The inner planet displaces the star
from the barycenter of the inner binary by an amount
\begin{equation}
x_0=-a_1\mu_1\sin{\left[2\pi(t-t_0)/P_1\right]}
\end{equation}
where the inner binary undergoes a transit at time $t_0$ and $P_1$
is the orbital period of the inner binary.  Thus, the timing deviation of
the $m$th transit of the outer planet is
\begin{equation}
\delta t_2 \approx -\frac{x_0}{v_2-v_0} \approx
-\frac{P_2a_1\mu_1\sin \left[2\pi(mP_2-t_0)/P_1\right]}{2\pi a_2}
\end{equation}
where $v_i$ is the
velocity of the $i$th body with respect to the line of sight.
Typically $v_0 \ll v_2$, so we have neglected $v_0$ in the second
expression in the previous equation.

The standard deviation of timing variations over many orbits is
\begin{equation}\label{sigcirc2}
\sigma_2 = \langle (\delta t_2)^2\rangle^{1/2} = \frac{P_2a_1\mu_1}{2^{3/2}\pi a_2}.
\end{equation}
Note that if the periods have a ratio $P_2$:$P_1$ of the form 
$q$:$1$ for some integer $q$, 
then the perturbations disappear
because the argument of the sine function is the same each orbit.

More interesting variations occur if either or both planets are
on eccentric orbits.  Because both planets are following Keplerian orbits, the transit timing variations and duration variations
can be computed by solving the Kepler problem for each
Jacobian coordinate.  Since we are assuming that the planets are coplanar
and edge-on, 4 coordinates for each planet suffice to determine the planetary
positions:  $e_{1,2}$, $a_{1,2}$, $\varpi_{1,2}$ (longitude of pericenter), 
and $f_{1,2}$ (true anomaly).  As in the circular case, the change in the transit timing is approximately $\delta t_2 \approx {x_0 / v_2}$.

The position of the star with respect to the barycenter
of the inner binary is
\begin{equation}
x_0 = -\mu_1 r_1 \sin{\left[f_1+\varpi_1\right]}.
\end{equation}
If $a_1\ll a_2$, the outer planet 
is in nearly the same position at the time of each transit 
and its velocity perpendicular to the line of sight is 
\begin{equation}
v_2={2\pi a_2\left(1+e_2\cos{\varpi_2}\right)\over P_2 \sqrt{1-e_2^2}}
\end{equation}
where we have used the fact that $f_2=-\varpi_2$ at the timing of
the transit.  Thus, to first order in $a_1/a_2$
\begin{equation}
\delta t_2 = -{P_2\mu_1 r_1 \sin{[f_1+\varpi_1]}\sqrt{1-e_2^2}\over
2\pi a_2(1+e_2\cos{\varpi_2})}.
\end{equation}

The standard deviation of $\delta t_2$ can be found analytically
as well.  To do this, I first calculate the mean transit deviation $\langle \delta t_2 \rangle$.  Over many transits by the outer planet, the inner binary's position
populates all of its orbit provided the planets do not have a period 
ratio that is the ratio of two integers.  Consequently, I find 
the mean transit 
deviation by averaging over the probability that the inner binary
is at any position in its orbit, $p(f_1)=n_1/\dot f_1$ (where $n_1=2\pi/P_1$),
times the transit deviation at that point.  This gives
\begin{eqnarray}
\langle \delta t_2 \rangle &=& {1 \over 2\pi} \int_0^{2\pi} df_1 \delta t_2
p(f_1) \cr
&=& -{3 \over 2} \mu_1 {a_1\over v_2} e_1 \sin{\varpi_1}.
\end{eqnarray}
The linear scaling with $e_1$ is because the star spends more time near apoapse which displaces the average position of the star, and hence the transit geometry, away from the barycenter of the system.
The symmetry of the orbit about $\varpi=0$ and $\pi$ explains the dependence
on $\sin{\varpi_1}$.

A similar calculation gives $\langle \delta t_2^2 \rangle $ and the resulting standard deviation is
\begin{eqnarray}\label{standarddeviation2}
\sigma_2 &=&\left(\langle\delta t_2^2\rangle-\langle\delta t_2\rangle^2\right)^{1/2}\cr
&=&{P_2a_1\mu_1 \sqrt{1-e_2^2}\over 2^{3/2}\pi a_2(1+e_2\cos{\varpi_2})}\left[1-{e_1^2\over 2}
\left(1+\cos^2{\varpi_1}\right)\right]^{1/2}.
\end{eqnarray}
This agrees with equation (\ref{sigcirc2})
in the limit $e_1 \rightarrow 0$.  Averaging again over $\varpi_1$ and
$\varpi_2$, gives
\begin{equation}
\langle\sigma_2\rangle_{\varpi_1,\varpi_2} =
{P_2a_1\mu_1 \left[1-{3\over 4}e_1^2\right]^{1/2}\over 2^{3/2}\pi a_2(1-e_2^2)^{1/4}}.
\end{equation}
Note that an eccentric inner orbit 
reduces $\sigma_2$ because the inner binary
spends more time near apoapse as the eccentricity increases which, in turn  
reduces the variation in the position of the star when 
averaged over time.  
As $e_1$ approaches unity for an orbit viewed along the
major axis, $\sigma_2$ reduces to zero because the minor axis approaches
zero and there is no resulting variation in the $x_0$ position.

\section{An exterior planet on a large eccentric orbit\label{largeecc}}

In this section I include planet-planet interactions to find 
the timing variations of an interior transiting planet---on an orbit with negligible eccentricity---that are caused by a perturbing planet that is on an 
eccentric orbit with a semi-major axis that is much larger than
that of a transiting planet.  In this limit, resonances are
not important and the ratio of the semi-major axes can be used
as a small parameter for a perturbation expansion.  A general
formula for this case has been derived by \cite{bor03}.
Here I present a
shorter derivation which clarifies the primary physical effects
for coplanar orbits.

The essential physics of this scenario is that the presence of the outer planet alters the effective mass, and thus the period, of the inner binary---the outer planet acts, in the lowest multipole, as a sphere surrounding the inner binary.  As the outer planet moves along its eccentric trajectory, its proximity to the inner binary changes.  The periodically changing distance between the perturbing planet and the inner binary causes the effective mass of the inner binary to change in tandem with the position of the outer planet.  So, as the outer planet moves inward, the inner binary slows in its orbit; as the outer planet moves outward, the inner binary orbits more rapidly.  These variations in the period of the inner binary translate to the timing deviations of the interior transiting planet.

The naive model of a pulsating sphere provides the proper scaling relations for the dispersion in the timing deviations.  I first derive this quantity using this approach since it encapsulates the relevant physics of the system.  Later, I provide a more detailed derivation of the transit timing variations that can be applied more generally.

Kepler's Laws state that the period of the inner binary scales as $P_1 \sim 1/\sqrt{G \rho_1}$ where $\rho_1 = m_0/a_1^3$ is roughly the average density of material inside the orbit of the transiting planet.  The outer planet, located a distance $r_2$ from the barycenter, acts to reduce this density, now denoted $\rho$, and alter the period of the inner binary.  The difference between the nominal period and the period $P$ that includes the effects of the outer planet is
\begin{equation}
\begin{split}
P_1 - P &\sim \frac{1}{\sqrt{G\rho_1}}\left(1 - \sqrt{\frac{\rho_1}{\rho}}\right) \\
&\simeq \frac{1}{\sqrt{G\rho_1}}\left(1- \left(\frac{m_0 r_2^3 - m_2 a_1^3}{m_0 r_2^3} \right)^{-1/2} \right)\\
&\simeq P_1\left(\frac{m_2}{m_0}\right)\left( \frac{a_1}{r_2} \right)^3.
\label{perioddiff}
\end{split}
\end{equation}
This result gives the timing deviation that results from one orbit of the transiting planet.  These deviations add over several orbits---roughly $P_2/P_1$ orbits---which serves to increase the typical timing deviation to an amount that is larger than that of a single orbit.  If we characterize the distance to the second planet as $r_2 \sim (1+e)a_2$ the resulting dispersion of timing deviations is
\begin{equation}
\sigma_1 \sim e_2 P_2 \left(\frac{m_2}{m_0}\right)\left(\frac{a_1}{a_2}\right)^3.
\label{scalesig2}
\end{equation}
I note that this result does not change if the nominal period $P_1$ in the derivation includes the average effect of the perturbing planet.

The item of fundamental importance is the variation in the distance $r_2$ in (\ref{perioddiff}) which result in the factor of $e_2$ in equation (\ref{scalesig2}).  The factor of $P_2$ encapsulates both the period of the transiting planet and the time scale over which the timing deviations accumulate, the mass ratio gives the relative strengths of the primary, central force which causes the orbit of the transiting planet and the secondary, perturbing force which causes the deviations in the orbit, and the factor of $e_2(a_1/a_2)^3$ characterizes the change in the volume of the orbit of the perturbing planet (volume being appropriate since the lowest multipole approximation is to treat the perturbing planet as a spherically symmetric change in the density of the system).

I now derive, in more detail, the individual transit timing variations as well as the dispersion in those deviations that occur in this scenario.  
The equations describing the inner binary can be divided into 
a Keplerian equation (\ref{jackepler}) and a perturbing force
proportional to $m_2$.  The perturbing acceleration $\delta \ddot{\bf r_1}$ 
on the inner binary due to the outer planet is given by
\begin{equation}
\delta \ddot{\bf r_1} = -GM\mu_2 {{\bf r}_1 - {\bf r}_{21}
\over \vert {\bf r}_1 - {\bf r}_{21}\vert^3}
- GM\mu_2 {{\bf r}_{21} \over r_{21}^3 }.
\end{equation}
We expand this in a Legendre series
and keep terms up to first order in the ratio of the radii
of the inner and outer orbit,
\begin{equation}
\delta\ddot{\bf r_1} = -{G m_2 \over r_2^3}\left[{\bf r}_1-3{{\bf r}_1\cdot
{\bf r}_2\over r_2^2}{\bf r}_2\right] + {\cal O}(r_1/r_2)^2.
\end{equation}

To compute the perturbed orbital period we must find
the change in the force on the inner binary due to the outer planet
averaged over the orbital period of the inner binary.
Since the inner binary is nearly circular, the angle of the
inner binary is given by  $\theta_1 = f_1 + \varpi_1 \simeq n_1(t-\tau_1) \varpi_1$, where I have approximated $e_1\simeq 0$.  
Differentiating this with respect to time gives
\begin{equation}
\dot \theta_1 = \dot n_1 (t-\tau_1) + n_1 - n_1\dot \tau_1.
\end{equation}
Now, we write $\dot n_1 = -3n_1/(2a_1)\dot a_1$, and
express $\dot a_1$, $\dot \varpi_1$,
and $\dot \tau_1$ in terms of the radial, tangential, and normal
components of the force \citep[see section 2.9 of][]{mur99}.  Plugging
these expressions into $\dot\theta_1$ gives a cancellation of most 
terms to lowest order in $e_1$, and after setting the normal force to 
zero leaves the remaining term
\begin{equation}
\dot \theta_1 = n_1 \left(1 - {2 a_1^2 \over G(m_0+m_1)}\bar R\right),
\end{equation}
where $\bar R$ is the radial disturbing force per unit mass, $ \bar R = (\delta
\ddot {\bf r}_1 \cdot \hat {\bf r}_1) = \frac{1}{2} Gm_2a_1/r_2^3$.

Thus, the change in the effective mass of the inner binary due to the presence of the second planet is actually $-{1 \over 2}m_2(a_1/r_2)^3$, 
which results in a slight increase in the period of the orbit.  
The increase in period would be constant if
the second planet were on a circular orbit.  However, for an eccentric orbit, 
the time variation of $r_2$ induces 
a periodic change in the orbital frequency of the inner binary
with period equal to $P_2$.

Now, the time of the $(N+1)$th transit occurs at 
\begin{eqnarray}\label{eclipsetime}
t_{ec}-t_0&=&\int_{f_0}^{f_0+2\pi N} df_1 \dot \theta_1^{-1} \cr
&=& \int_{f_0}^{f_0+2\pi N} df_1
n_1^{-1} \left[1+{m_2 \over m_0+m_1} \left({a_1\over r_2}\right)^3\right].
\end{eqnarray}
where $f_0$ is the true anomaly of the inner binary at the time of the first 
transit.  Following \cite{bor03}, the variable of
integration can be changed from $f_1$ to $f_2$, the true anomaly of the outer planet,
\begin{equation}
df_1 = {P_2\over P_1}{r_2^2\over a_2^2(1-e_2^2)^{1/2}}df_2.
\end{equation}
Since we are assuming that the orbit of the outer planet is eccentric, 
$r_2=a_2(1-e_2^2)/(1+e_2\cos{f_2})$, which 
gives the transit time
\begin{equation}\label{eccentricouter}
t_{ec}-t_0=NP_1+{m_2\left(1-e_2^2\right)^{-3/2}\over 2\pi(m_0+m_1)}{P_1^2\over P_2}
 \left(f_2+e_2\sin{f_2}\right),
\end{equation}
where $f_2$ is the true anomaly of the outer binary at the
timing of the $(N-1)$ transit.  The unperturbed $f_2$ includes 
the mean motion, $n_2(t-\tau_2)$, which grows linearly with time.
To find the deviation of the
time of transits from a uniform period, I subtract off this mean
motion as well as $NP_1$ which results in 
\begin{eqnarray}
\delta t_1 &=& \beta
\left(1-e_2^2\right)^{-3/2} \left[f_2-n_2(t-\tau_2)+e_2\sin{f_2}\right]\cr
\beta&=& {m_2 \over 2\pi(m_0+m_1)}{P_1^2\over P_2}.
\end{eqnarray}
This agrees with the expression of \cite{bor03} in the limit $I=1$
(i.e. coplanar orbits).

\begin{figure}
\includegraphics[width=\fsize]{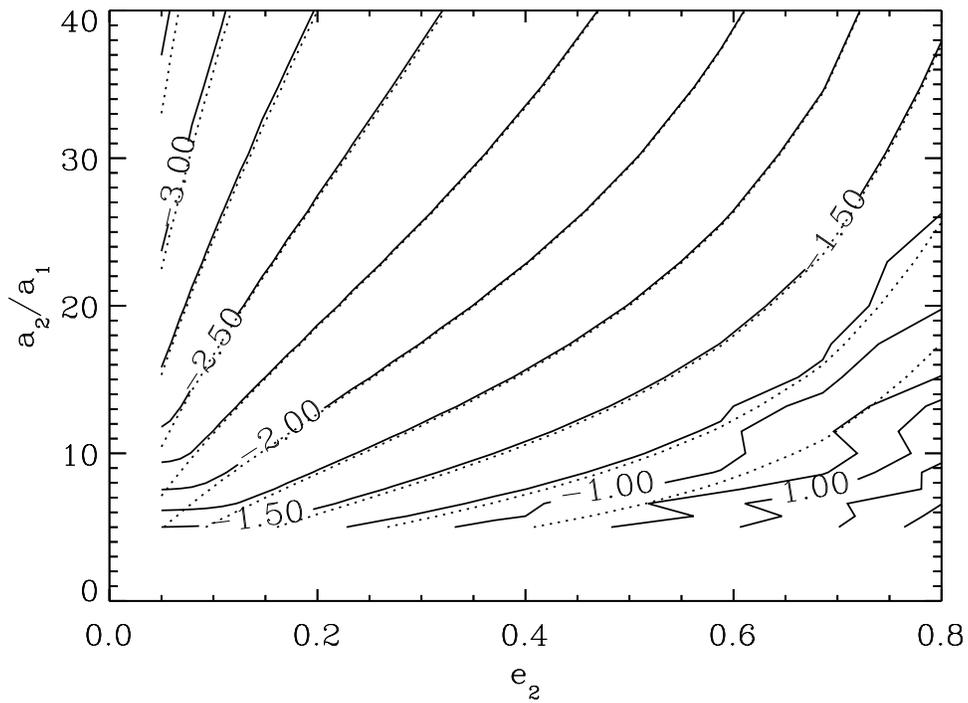}
\caption{Contour plot of the logarithm of the dispersion of the normalized 
timing variations, $\log{(\sigma_1 n_1 \mu_2^{-1})}$, for an inner circular 
planet and an outer eccentric planet (for example, at the -2.00 contour an
orbit lasting $2\pi$ years with a perturbing planet of mass $10^{-3}
{\rm M_\odot}$
would have a transit timing standard deviation of $10^{-5}$ years, or 
5 minutes). The dotted line is the approximation given in equation 
(\ref{sig2approx}).}
\label{fig2}
\end{figure}

Numerical calculation of the 3-body
problem show that this approximation works extremely well in the 
limit $r_1\ll r_2$ (see Figure \ref{fig2}).  
If $P_1\ll P_2$ and the period ratio 
is non-rational, then over a long time the transits of the inner planet 
sample the entire phase of the outer planet.  Thus, we can find the
standard deviation of the transit timing variations as in equation 
(\ref{standarddeviation2})
\begin{equation}
\sigma_1 = \langle \delta t_1^2 \rangle^{1/2} = {1 \over 2\pi} \int_0^{2\pi} df_2 \delta t_1^2 p(f_2),
\end{equation}
since $\langle \delta t_1\rangle=0$, where $p(f_2)=n_2/\dot f_2$.
This integral is intractable analytically, but
an expansion in $e_2$ yields
\begin{equation}\label{sig2approx}
\sigma_1 =  {3\beta e_2 \over \sqrt{2}\left(1-e_2^2\right)^{3\over 2}}
\left[1-{3\over 16}e_2^2-{47\over 1296}e_2^4-{413\over 27648}e_2^6\right]^{1/2},
\end{equation}
which is accurate to better than 2 per cent for all $e_2$ and agrees with equation (\ref{scalesig2}) in the limit as $e_2 \rightarrow 0$.
Figure \ref{fig2} shows a comparison of this approximation with the
exact numerical results averaged over $\varpi_2$ (since there is a
slight dependence on the value of $\varpi_2$).  This approximation
breaks down for $a_2(1-e_2) \lesssim 5 a_1$ since resonances and higher
order terms contribute strongly when the planets have a close approach.
It also breaks down for $e_2 \lesssim 0.05$ since the perturbations to
the semi-major axes caused by interactions of the planets contribute
more strongly than the tidal terms which become weaker with smaller
eccentricity.

\section{Non resonant planets on initially circular orbits\label{nonresonant}}
\label{circularpert}

In this section I calculate the amplitude of timing variations for
two planets on nearly circular orbits.  That is, on orbits where a first-order, perturbative expansion of the planetary motion is valid.  
The resonant forcing terms 
are most important factors that determine the amplitude of the periodic timing variations, even for non-resonant planets.  
The planets interact most strongly at conjunction, so the
perturbing planet causes a radial kick to the transiting planet 
inducing eccentricity into its orbit.  This change in eccentricity, in turn, causes a change in the semi-major axis by an amount given by the Tisserand relation
\begin{equation}
\frac{da}{de} = \frac{2a^{5/2}e}{a^{3/2}-1}
\end{equation}
where $a$ is, in this case, the ratio of the semi-major axes of the planets.  

Both the change in eccentricity and the change in semi-major axis contribute to the variations in the period of each planet.  This can be seen by Taylor-expanding the time-dependent angular velocity of the planet $\dot{\theta}$ to first order in $e$
\begin{equation}
\dot\theta = {n \left(1+e \cos{f}\right)^2 \over 
\left(1-e^2\right)^{3/2}} 
\approx n_{0} + \delta n + 2 e n_{0}\cos{\left[\lambda-\varpi\right]},
\label{thetadot}
\end{equation}
where $n_0$ is the nominal mean-motion of the planet, $\delta n$ reflects the variations caused by changes to the mean-motion, and the oscillatory term of amplitude proportional to $e$ quantifies the variations caused by changes to the eccentricity.  I first derive the timing deviations for when the change in mean-motion dominates the transit time variations.  Then, I derive an approximation to the variations that result when the induced eccentricity dominates the timing deviations---which happens to occur farther from the resonances than the mean-motion dominated variations.  Finally, I derive in detail the eccentricity dominated variations using perturbation theory.  I show the variations that occur when the planets are in mean-motion resonance in the next section.

\subsection{Mean-motion dominated variations}

As previously stated, the radial perturbing force on the transiting planet is largest when the planets are in conjunction.  Since, in this limit where changes to the mean motion dominates the timing variations, the
planets are not exactly on resonance, the longitude of conjunction
will drift with time.  Eventually, the effects of the radial kicks cancel after the longitude
drifts by $\simeq \pi$ in the inertial frame.  Thus, the total amplitude
of the eccentricity grows over a time equal to half of the period of 
circulation of the longitude of conjunction. The closer the planets are 
to a resonance, the longer the period of circulation and thus the 
larger the eccentricity grows.  The change in eccentricity
causes a change in the semi-major axis and 
mean motion---the effect we study here.  Identifying the regime where the timing deviations are dominated by the mean motion will be easiest to identify once both the mean-motion effects and the eccentricity effects have been independently quantified and I defer to the end of the section \ref{eccdominate} for the location of this transition.

For two planets that are on circular orbits near a $j$:$j+1$ resonance,
conjunctions occur every $P_{conj}=2\pi/(n_1-n_2)\simeq jP$ (I take
the limit of large $j$ and we ignore factors of order unity).  Let us define 
the fractional distance from resonance, $\epsilon
= \vert1-(1+j^{-1})P_1/P_2\vert < 1$, 
where $\epsilon=0$ indicates exact resonance.  The longitude of conjunction 
changes with each successive conjunction and ultimately 
returns to its initial value over a period 
$P_{cyc}=Pj^{-1}\epsilon^{-1}$.  The number of conjunctions per cycle 
is $N_c=P_{cyc}/P_{conj}\simeq j^{-2}\epsilon^{-1}$.
Each conjunction changes the eccentricity of the planets by
$\Delta e \sim \mu_{pert} (a/\Delta a)^2$ (using the perturbation equations
for eccentricity and the impulse approximation, where $\mu_{pert}$
is the planet-star mass ratio of the perturbing planet).  Over half a cycle the
eccentricities grow to about $N_c \Delta e \sim \mu_{pert} (1-P_1/P_2)^{-1}
(j\epsilon)^{-1} \simeq \mu_{pert} \epsilon^{-1}$.  

To find the change in the transit timing that is caused by the $\delta n$ term in equation \ref{thetadot} I apply the Tisserand relation to the lighter planet (now using subscripts ``light" and ``heavy"), resulting in 
$\delta n_{light}/n_{light} \simeq j (\Delta e_{light})^2
\simeq j\mu^2 \epsilon^{-2}$ (where $\mu$ is $\mu_{heavy}$).  
These changes to the period accumulate over an entire cycle, giving
\begin{equation}\label{lightplanet}
\delta t_{light} \simeq \mu^2 \epsilon^{-3} P.
\end{equation}
By conservation of
energy, the fractional change in semi-major axis (or period) of the 
heavy planet is reduced by
a factor of $m_{light}/m_{heavy}$, so that 
\begin{equation}\label{heavyplanet}
\delta t_{heavy}\simeq (m_{light}/m_{heavy}) \mu^2 \epsilon^{-3} P.
\end{equation}

\subsection{Eccentricity dominated variations\label{eccdominate}}

For timing variations that are caused by changes to the eccentricity of the transiting planet we look at the eccentricity dominated term in (\ref{thetadot}).  Following the derivation from the previous section (up to the point where the tisserand relation is applied) we find that the eccentricity term gives a timing variation of 
\begin{equation}\label{wings}
\delta t \simeq \mu_{pert} \epsilon^{-1} P.
\end{equation}
This means that for $\epsilon > \mu^{1/2}$ 
the perturbed eccentricity dominates the timing deviations while closer to resonance for 
$j^{1/3}\mu^{2/3} < \epsilon < \mu^{1/2}$ the perturbed mean motion dominates 
(this range is the same for both the light and heavy planets, except for
factors of order unity).  For
smaller values of $\epsilon$, the planets are trapped in mean-motion resonance,
which was discussed in the previous section.   Half way between
resonances, $\epsilon\simeq j^{-2}$,  so the timing deviation become
\begin{equation}\label{halfres}
\delta t \sim 0.7 \mu_{pert} (a/(a_2-a_1))^2P.
\end{equation}
A more precise derivation in the eccentricity-dominated case
using perturbation theory is given in the next section.

\subsection{Perturbation theory derivation of eccentricity dominated variations}

In this section we consider more carefully the case of two planets
whose orbits are nearly circular and whose timing variations
are dominated by changes in eccentricity (see \S \ref{circularpert}).
The timing variations can
be computed from a Hamiltonian as described in \cite{mal93a}.  
We keep terms which are first order in the eccentricity because
mutual perturbations between the planets induces an eccentricity of
order $m_2/m_0$.  To first order in the eccentricities, the
Taylor-expanded Hamiltonian is\footnote{We have corrected equation
(26) in \cite{mal93a} which should have a $-\alpha\cos{\psi}$ in the
second line.}
\begin{eqnarray}
H_{int} =& -{Gm_1m_2 \over 2a_2} \left[2P-2\alpha \cos{\psi}
-e_1c^+_0\cos{\left(\lambda_1-\varpi_1\right)}\right.\cr
+&e_2d^+_0 \cos{\left(\lambda_2-\varpi_2\right)}\cr
-&e_1 \sum_{j=1}^\infty \left(c^-_j+\alpha\delta_{j,1}\right)\cos\left((j+1)\lambda_1-j\lambda_2-\varpi_1\right)\cr
-&e_1 \sum_{j=1}^\infty \left(c^+_{j}-3\alpha\delta_{j,1}\right)\cos\left((j-1)\lambda_1-j\lambda_2+\varpi_1\right)\cr
+&e_2 \sum_{j=1}^\infty d^-_j\cos\left(j\lambda_1-(j-1)\lambda_2-\varpi_2\right)\cr
+&\left.e_2 \sum_{j=1}^\infty \left(d^+_j-4\alpha\delta_{j,1}\right)\cos\left(j\lambda_1-(j+1)\lambda_2+\varpi_2\right)\right],
\end{eqnarray}
where $\delta_{i,j}$ is the Kronecker delta, $c^{\pm}_j=\partial_\alpha b^{(j)}_{1/2}\pm 2jb^{(j)}_{1/2}$, $d^{\pm}_j=c^{\pm}_j+b^{(j)}_{1/2}(\alpha)$,
$\alpha = a_1/a_2$, $b^{(j)}_{1/2}(\alpha)$ is the Laplace coefficient,
\begin{equation}
b^{(j)}_{1/2}(\alpha)= {1 \over \pi} \int_0^{2\pi} d\theta {\cos{j\theta} \over
\sqrt{1-2\alpha\cos{\theta}+\alpha^2}},
\end{equation}
where $\psi=\lambda_1-\lambda_2$,
\begin{equation}
P=P(\psi,\alpha)=\left(1-2\alpha\cos{\psi}+\alpha^2\right)^{-1/2},
\end{equation}
and 
\begin{equation}
\partial_{\alpha} b^{(j)}_{1/2}(\alpha) \equiv \alpha{\partial \over \partial \alpha}b^{(j)}_{1/2}(\alpha).
\end{equation}
This equation includes no secular terms since these are higher order
in the eccentricity.  Note that since we have only included the first order 
terms in the eccentricity, the resonant arguments which appear have ratios 
$j+1$:$j$ and $j$:$j+1$ for the mean longitudes. 

The perturbed semi-major axis is given in \cite{mal93b}, and 
we compute the perturbed eccentricity and longitude of periastron
using $h_1=e_1\sin{\varpi_1}$ and 
$k_1=e_1\cos{\varpi_1}$.  Keeping all the resonance terms that exist to first 
order in the eccentricities gives the equations of motion for $h_1, k_1$,
\begin{eqnarray}
\dot h_1 &=& -{1 \over 2} n_1 \alpha {m_2 \over m_0} 
\left[c^+_0 \cos{\lambda_1} \right.\cr
&+& \left. \sum_{j=1}^{\infty} \left\{\left(c^+_j -3\alpha\delta_{j,1}\right)\cos{[(j-1)\lambda_1-j\lambda_2]} \right.\right.\cr
&+&\left.\left.\left(c^{-}_j +\alpha\delta_{j,1}\right)\cos{[(j+1)\lambda_1-j\lambda_2]}\right\}\right],\cr
\dot k_1 &=& {1 \over 2} n_1 \alpha {m_2 \over m_0} 
\left[c^+_0 \sin{\lambda_1} \right.\cr
&-&
\sum_{j=1}^{\infty} \left\{\left(c^+_j-3\alpha\delta_{j,1}\right)\sin{[(j-1)\lambda_1-j\lambda_2]} \right.\cr
&-&\left.\left.\left(c^{-}_j+\alpha\delta_{j,1}\right) \sin{[(j+1)\lambda_1-j\lambda_2]}\right\}\right].\cr
\end{eqnarray}

To find the change in the transit timing we use the orbital elements
to compute the variation in $\dot\theta_1$.  To first order in $e_1$
\begin{equation}
{\dot\theta_1 \over n_{10}}\approx 1 + {\delta n_1\over n_{10}} + 2 k_1 \cos{\left[n_{10}t+\lambda_{10}\right]}+2h_1\sin{\left[n_{10}t+\lambda_{10}\right]}.
\end{equation}
Since we begin with zero eccentricity, we ignore perturbations to $\lambda$ in 
the $\sin$ and $\cos$ terms in this equation.
As in equation (\ref{eclipsetime})
where $\delta \dot\theta_1 = \dot\theta_1 -n_{10}$, we 
integrate this equation to find
\begin{eqnarray}\label{sigcirc1}
\delta t_1 =& 3{m_2 \over m_0}\alpha {n_1 \over \left(n_1-n_2\right)^2} \left(Q(\psi)-{2\psi K(k) \over \pi(1+\alpha)}-\alpha \sin{\psi}\right)\cr
-& n_1{m_2 \over m_0}\alpha \left[{c^+_0 \over n_1^2}\sin{(\lambda_1-\lambda_{10})}\right.\cr
+&\sum_{j=1}^\infty {c_j^+ -3\alpha\delta_{j,1} \over (j-1)n_1-jn_2} \left({\sin{[j\psi]} \over j(n_1-n_2)}-{\sin{[n_1t+j\psi_0]}\over n_1}\right)\cr
-&\left.\sum_{j=1}^\infty {c_{j}^- + \alpha\delta_{j,1} \over (j+1)n_1-jn_2} \left({\sin{[j\psi]} \over j(n_1-n_2)}-{\sin{[n_1t-j\psi_0]}\over n_1}\right)\right],
\end{eqnarray}
where $n_1$ and $n_2$ are taken at their initial values,
$\lambda_{10}=\lambda_1(t=0)$, $\lambda_{20}=\lambda_2(t=0)$, 
$k=2\sqrt{\alpha}/(1+\alpha)$, $K(k)$ is the complete elliptic integral, 
$Q(\psi)$ is defined in the appendix of \cite{mal93b},
and we have dropped any terms which vary linearly with time.  

\begin{figure}
\includegraphics[width=\fsize]{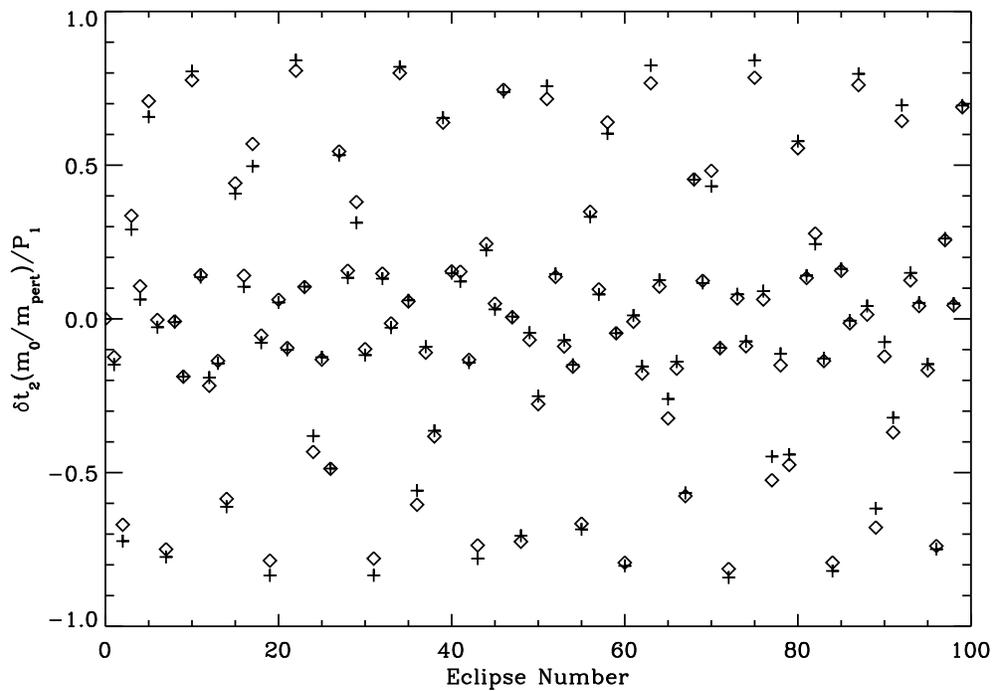}
\caption{Deviations from uniform times between transits for
a transiting outer planet with semi-major axis ratio 1.8. Crosses
represent the analytic result, equation (\ref{sigcircjason}), while
the diamonds are from a numerical integration of the equations
of motion.  The horizontal axis is the number of the transit, while 
the vertical axis shows the timing differences in units of $P_1$,
the period of the inner perturbing planet, multiplied by $m_0/m_{pert}$.}
\label{fig8}
\end{figure}

A similar calculation can be carried
out for perturbations by a planet interior to the transiting planet,
\begin{eqnarray}\label{sigcircjason}
\delta t_2 =& -3 \frac{m_1}{m_0} \frac{n_2}{(n_1-n_2)^2} \left( Q(\psi) - \frac{2\psi K(k)}{\pi(1+\alpha)} - \alpha \sin(\psi) \right) \cr
+& n_2 \frac{m_1}{m_0} \left[\frac{d_0^+}{n_2^2}\sin[\lambda_2-\lambda_{20}] \right. \cr
+& \sum_{j=1}^{\infty} \frac{d_j^+ - 4\alpha\delta_{j,1}}{(jn_1-(j+1)n_2)} \left( \frac{\sin[j\psi]}{j(n_1-n_2)} - \frac{\sin[n_2t+j\psi_0]}{n_2} \right) \cr
-&\left. \sum_{j=1}^{\infty} \frac{d_j^-}{(jn_1-(j-1)n_2)} \left( \frac{\sin[j\psi]}{j(n_1-n_2)} - \frac{\sin[n_2t-j\psi_0]}{n_2} \right) \right]
\end{eqnarray}
A comparison of these equations to numerical calculations is shown in Figure 
\ref{fig8}.  The planets are on initially 
circular orbits with a semi-major axis ratio of 1.8, or period ratio of 2.4.  
The masses are equal and the planets start aligned along the line of sight at the 
first transit.  

In the case that the semi-major axis dominates the timing variations,
one can take perturbed value of the eccentricity ($h$ and $k$) and
compute the change in semi-major axis due to these eccentricities.
The result is a double-series over resonant terms, so I do not
reproduce it here.

So far we have discussed the timing variations for planets near, but not in 
a first order resonance.  For larger period ratios, the eccentricity
of the inner planet grows to $e_{in} \simeq \mu_{out}(P_{in}/P_{out})^2$,
so $\delta t_{in} \sim \mu_{out} (P_{in}/P_{out})^2 P_{in}$.
For an outer transiting planet the motion of the star dominates
over the perturbation due to the inner planet for 
$P_{out}>(2\pi)^{3/4}P_{in}$.

\begin{figure}
\includegraphics[width=\fsize]{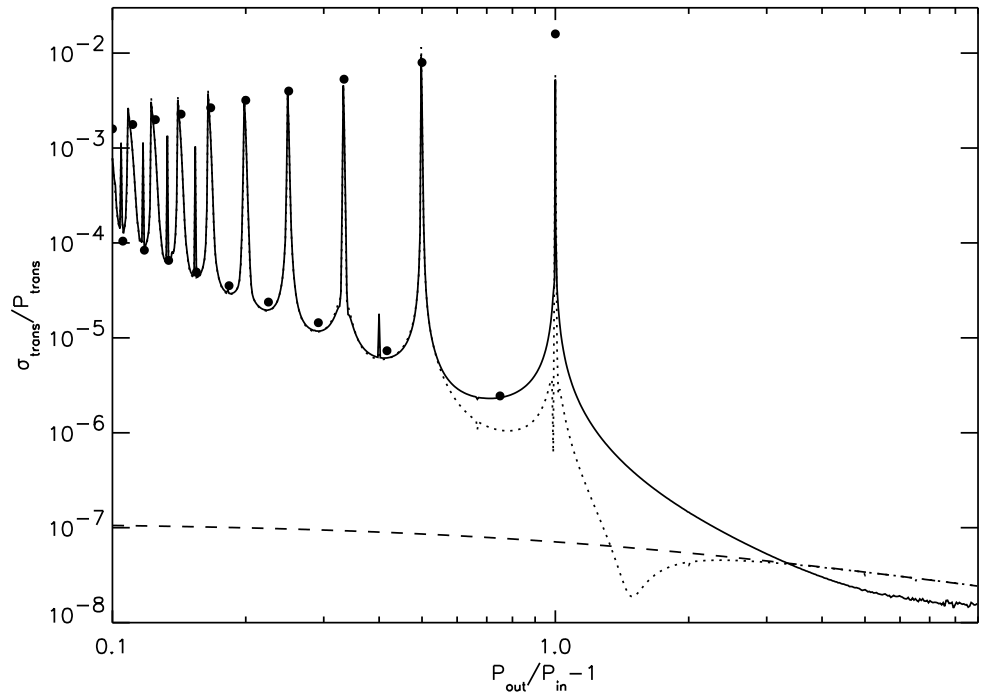}
\caption{Transit timing standard deviation for two planets of mass
$m_{trans}=10^{-5}m_0$ and $m_{pert}=10^{-6}m_0$ on initially
circular orbits in units of the period of the transiting planet.  The 
solid (dotted) line is the numerical calculation for the inner (outer)
planet averaged over 100 orbits of the outer planet with the planets 
initially aligned with the observer.  
The dashed line is equation (\ref{sigcirc2}).  The large dots are
equations (\ref{reemeqn}) on resonance and equation (\ref{halfres})
halfway between resonances.}
\label{fig3}
\end{figure}

Figure \ref{fig3} shows a numerical calculation of the
standard deviation of the transit timing variations.  I 
used small masses to avoid chaotic behavior since resonant
overlap occurs for $j\gtrsim \mu^{-2/7}$ \citep{wis80}.  Figure \ref{fig4}
zooms in on the 2:1 resonance.  As predicted, the amplitude scales 
as $\epsilon^{-1}$ (equation \ref{wings}), and then steepens to $\epsilon^{-3}$ 
(equations \ref{heavyplanet} and \ref{lightplanet}) closer to 
resonance.  Since the strength of the perturbation is independent
of whether the perturbing planet is interior or exterior,
the strength of the resonances are similar and the shape of the
standard deviation of the transit timing variations
is symmetric about $P_{in}=P_{out}$.
The dashed curve in Figure \ref{fig3} shows the analytic approximation
from equation (\ref{sigcirc2}), which agrees well for $P_{pert}<(2\pi)^{-3/4} P_{trans}$.
The numerical results match the perturbation calculation, equations
(\ref{sigcirc1}) and (\ref{sigcircjason}),
except for near resonance where the change in mean-motion dominates (we have
not bothered to overplot the perturbation calculation since it is
indistinguishable from the numerical results).

There is a dip in $\sigma_2$ near $P_{out}=2.5 P_{in}$ which 
occurs because the amplitude of the timing differences due to
the orbit of the star about the barycenter (eqn. \ref{sigcirc2}) are 
opposite in sign and
comparable in amplitude to the differences due to the perturbation of
the outer planet by the inner planet (eqn. \ref{sigcircjason}).

\section{Planets in resonance\label{resonant}}

The results of the previous section break down near each mean-motion resonance because linear perturbation theory, which assumes that the perturbations are small, fails in the resonant regime.  We must consider the effects of higher order terms and changes to the orbital elements of the perturbing planet in order to understand the effects of resonance.  I present the case of low, initially
zero, eccentricity where the standard analyses of this case
\citep[e.g.][]{mur99} to be incorrect.  Here I provide a physically motivated, order
of magnitude, derivation of the perturbations and the transit timing
variations for two planets in a first-order mean-motion resonance.
A rigorous derivation is left for elsewhere, but 
I verify these findings with numerical simulations.

Consider a first order, $j$:$j$+$1$, resonance where the lighter
planet is a test particle.  Qualitatively, the physics of low
eccentricity resonance is as follows: on the nominal resonance, the
two planets have successive conjunctions at exactly the same longitude
in inertial space. The strong interactions that occur at conjunctions 
build up the eccentricity of the test particle and cause a change in semimajor
axis and period. The change in period of the test particle 
causes the longitude of conjunction to drift. Once the
longitude of conjunction shifts by about $\pi$ relative to the
original direction, the eccentricity begins to decrease making a
libration cycle. The libration of the semi-major axes causes
the timing of the transits to change.  Note that the libration of the longitude of conjunction distinguishes the resonant interaction with the non-resonant cases presented earlier.  In non-resonant systems the longitude of conjunction constantly drifts in the same direction.

The above, qualitative discussion leads directly to an estimate of the
drifts in transit times. Within each libration cycle the longitude of
conjunction shifts by about half an orbit, mostly due to the period
change of the lighter planet. Since conjunctions occur only once every
$j$ orbits the largest transit time deviation of the lighter planet during
the period of libration is $P/j$
(in this order of magnitude derivation we ignore factors of order 
unity, and take the limit 
of large $j$ so that $j\simeq j+1$ and $P_2\simeq P_1$). 
The
observationally more interesting case is probably that in which the
heavier planet is the transiting one. 
Then, conservation of energy
for the orbiting planets implies that the change in periods is inversely
proportional to the masses, therefore the timing variations are given by
$(m_{light}/m_{heavy})P/j$.  We find an excellent fit to the data for
\begin{equation} \label{reemeqn}
\delta t_{max} \sim {P\over 4.5j}{m_{pert}\over m_{pert}+m_{trans}}.
\end{equation}
The calculations shown in Figure \ref{fig3} verify this analytic scaling 
with $j$.
  
Calculating the libration period is a little more
complicated, but still straightforward. Suppose the period of the test
particle deviates from the nominal resonance by a small fraction
$\epsilon$. Then, consecutive conjunctions drift in longitude by about 
$2\pi j^2 \epsilon$. The number of conjunctions, $N_c$,
before a drift of order $\pi$ in the longitude of conjunctions
accumulates is $N_c \sim j^{-2}\epsilon^{-1}$.  We now estimate $\epsilon$
indirectly.  The test particle gains an eccentricity of order $j^2\mu$ in
each conjunction due to the radial force from the massive planet (this
can be computed from the impulse approximation and the perturbation
equation for eccentricity).  The eccentricity given 
in $N_c$ conjunctions is then of order $\Delta e \sim \mu \epsilon^{-1}$.  
Using the Tisserand relation, the fractional change in semimajor axis associated 
with this change in eccentricity is $j \mu^2 \epsilon^{-2}$.  
Since this is also the fractional change in the period
we have $\epsilon \sim j^{1/3} \mu^{2/3}$ and a libration period of
\begin{equation}\label{plib}
P_{lib} \sim 0.5 j^{-1} \epsilon^{-1} P \sim 0.5 j^{-4/3} \mu^{-2/3}P.
\end{equation}

\begin{figure}
\includegraphics[width=\fsize]{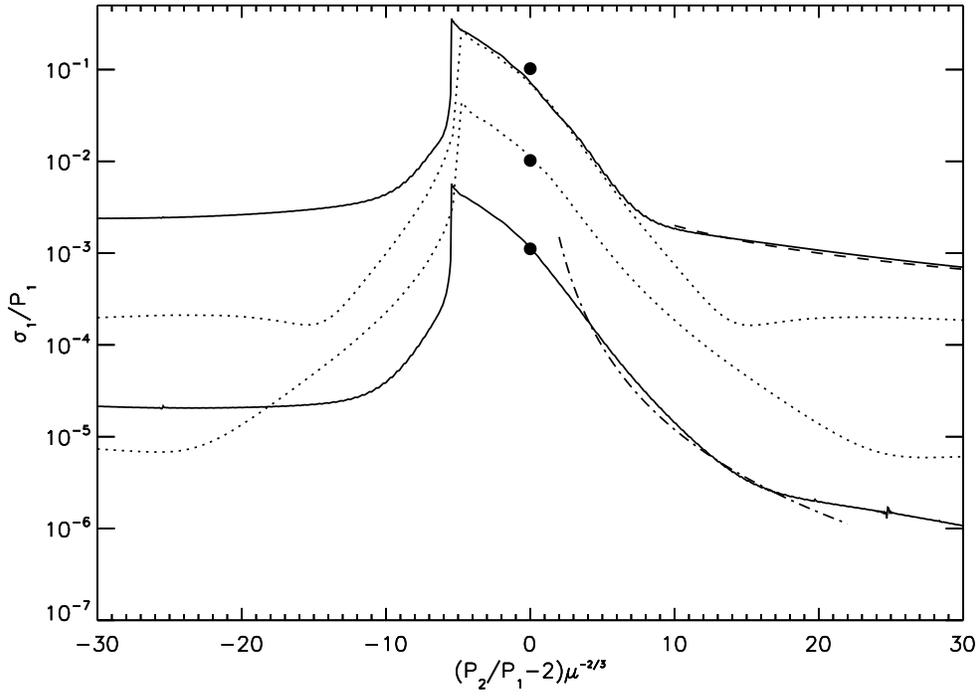}
\caption{Amplitude near the 2:1 resonance versus the difference in 
period from exact resonance for two systems:  one with
$m_1=10^{-5}m_0$ (top solid) and $m_2=10^{-3}$
(lower solid), and the other with $m_1=10^{-6}m_0$ (top dotted) and $m_2=10^{-5}$ 
(lower dotted).  The large dots are equation (\ref{reemeqn}).
The dashed line is equation (\ref{wings}), while the dash-dot line is 
equation (\ref{heavyplanet}).}
\label{fig4}
\end{figure}

We numerically computed the amplitude and period of the transit timing 
variations at the 2:1 resonance.  Figure \ref{fig4} shows a plot of the 
amplitude of the timing variations versus the mass ratio of the
perturbing planet to the transiting planet.  As predicted, the amplitude is of
order the period of the transiting planet when the transiting planet is lighter, 
and varies as the mass ratio when the transiting planet is heavier. 
The libration period measured from the numerical simulations shows the 
predicted behavior, scaling precisely
as $\mu^{-2/3}$ for the more massive planet (with a coefficient of 
$\sim 0.7$ for $j=1$ and $0.5$  for $j>1$ in equation \ref{plib}).
We have compared the numerical values of the amplitude
and period of libration on resonance as a function of $j$.  Despite the
fact that the above scalings were derived in the large-$j$ limit, the
agreement is better than 10 per cent for $j\ge 2$, and accurate to about 40
per cent for $j=1$.

Figure \ref{fig4} shows the more detailed
behavior of the amplitude of the dispersion of the timing deviations near the 2:1 resonance.  The amplitude is
maximum slightly below resonance at the location of the cusp.  This may
be understood as follows:  since the simulations are started with
$e_1=e_2=0$, after conjunction the eccentricity grows and the outer planet
moves outwards, while the inner planet moves inward.  This causes
the planets to move closer to resonance, causing a longer time between
conjunctions, leading to a larger change in eccentricity and semi-major
axis.  The cusp is the location where the planets reach exact resonance
at the turning point of libration, at which point $\delta t$ is maximum.
To the right of the cusp, the libration causes the planets to overshoot
the resonance, so the change in eccentricity and semi-major axis is 
somewhat smaller, and hence the amplitude is smaller.
Figure \ref{fig4} shows that the width of the resonance scales as
$\mu^{2/3}$ (the horizontal axis has been scaled with $\mu^{-2/3}$ so
that the curves overlap), so for larger mass planets
the resonant variations have a wider range of influence than the non-resonant
variations discussed in the previous section.  The curves in Figure \ref{fig4}
demonstrate that on-resonance the amplitude scales as $min(1,\mu_{pert}/
\mu_{trans})/j$, while off-resonance the amplitude scales as $\mu_{pert}$.

\section{Non-zero eccentricities\label{generalorbs}}

\begin{figure}
\includegraphics[width=\fsize]{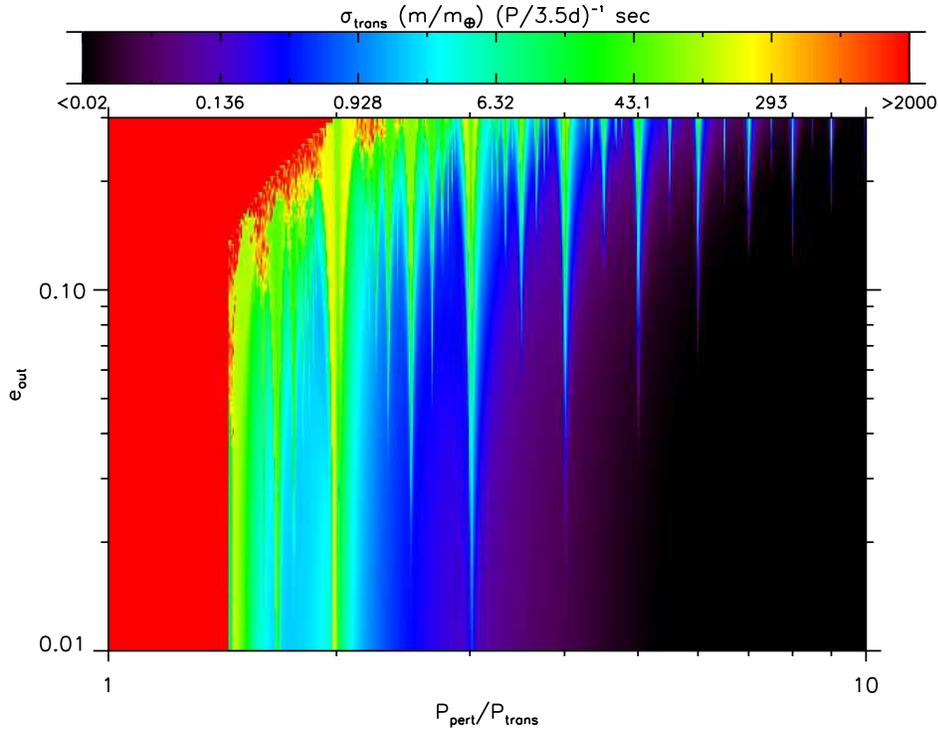}
\caption{Dispersion, $\sigma_{trans}$, of timing variations of HD209458b due to 
perturbations induced by an exterior earth-mass planet with various 
eccentricities and periods.  The color bar has $\sigma_{trans}$ for a
planet of mass $m_\oplus$ and for a transiting period
of 3.5 days.  Near the period ratio of 4:3 the 
system becomes chaotic.  The increase in the number of resonance peaks, 
particularly those beyond the 2:1 resonance, are from higher order 
terms in the expansion of the Hamiltonian which were truncated for
the near-circular case.}
\label{fig5}
\end{figure}

When either eccentricity is large enough, higher order resonances
become important.  In particular, the resonances that are 1:$m$ begin
to dominate as the ratio of the semi-major axes becomes large; as
the eccentricity of the outer planet approaches unity these resonances
become as strong as first order resonances \citep{pan04}.  Figure
\ref{fig5} shows the results of a numerical calculation where the
transiting planet HD209458b, with a mass of approximately 0.67 Jupiter
masses, is perturbed by a $1{\rm M_\oplus}$ planet with various eccentricities
(we have taken HD209458b to have a circular orbit).
Near the mean-motion resonances the signal is large enough that an 
earth-mass planet would be detectable with current technology. The 
amplitude increases everywhere with eccentricity.  This graph can 
be applied to systems with other masses and periods as the timing 
variation scales as $\delta t \propto P_{trans} m_{pert}$ (except
for planets trapped in resonance).

When both planets have non-zero eccentricity, the parameter
space becomes quite large:  the 4 phase space coordinates for each planet,
assuming both are edge-on and 2 mass-ratios give 10 free parameters.
On resonance, the analysis remains similar to the circular case.
The libration amplitude will still be of order $\sim Pj^{-1}$ for the 
lighter planet and $\sim Pj^{-1}(\mu_{light}/\mu_{heavy})$ 
for the heavier planet. However, 
the period of libration will decrease significantly as the eccentricity
increases since $P_{lib} \simeq e^{-1/2} \mu^{-1/2}$.

On the secular time-scale, the precession of the orbits will lead to a
significant variation in the transit timing \citep{mir02}.  
The period of precession, $P_\nu$, may be driven
by other planets, by general relativistic effects, or by a non-spherical
stellar potential, but leads to a magnitude of transit timing
deviation which just depends on the eccentricity for $P_\nu \gg P$.
\cite{mir02} showed that the maximum deviation for $e \ll 1$
is given by
\begin{equation}\label{maxprecession}
\delta t = \frac{eP}{\pi}
\end{equation}
and the timing variations vary with a period that is equal to the period of precession.
For arbitrary eccentricity, the maximum deviation is
\begin{equation}
\delta t = {P \over 2 \pi}\left[\sin^{-1}{y}
+\sin^{-1}{z} +\sqrt{2x-x^2-x^4}\right],
\end{equation}
where $x=(1-e^2)^{1/4}$, $y=(1-x)/e$, and $z=(1-x^3)/e$ (this is derived from the Keplerian solution with a
slowly varying $\varpi$).  This 
approaches $P/2$ as $e\rightarrow 1$.
Typically the eccentricity will vary on the secular time-scale,
so these expressions only hold as long as the variation in $e$ is
much smaller than its mean value.

Rather than systematically studying the entire parameter
space, we now investigate several specific cases of known extrasolar 
planets to demonstrate that detection of this effect should be possible 
once a transiting multi-planet system is found.  Most of these systems
have non-zero eccentricities and several are in resonance, causing
a significant signal.  We summarize the amplitude of the signals
of most known multi-planet systems, if they were seen edge-on, in 
Table 1 (in some cases 
other planets are present which would cause additional perturbations).

\begin{table}
\caption{Timing variations for known multi-planet systems}\label{tab1}
\begin{tabular}{@{}lcccc}
System & $P_{in}$ (d) & $P_{out}/P_{in}$& $\sigma_1$ & $\sigma_2$\cr
\hline
55 Cnc e, b      &  2.81  &  5.21  &  10.5 s  &  2.68 s  \cr
55 Cnc b, c      &  14.7  &  3.02  &  1.61 h  &  14.7 h  \cr
Ups And b, c     &  4.62  &  52.3  &  1.30 s  &  1.61 min  \cr
Gliese 876       &  30.1  &  2.027 &  1.87 d  &  14.6 h  \cr
HD 74156         &  51.6  &  39.2  &  4.98 min  &  42.4 min  \cr
HD 168443        &  58.1  &  29.9  &  12.9 min  &  2.62 h  \cr
HD 37124         &  152   &  9.81  &  3.43 d  &  11.2 d  \cr
HD 82943         &  222   &  2.00  &  34.9 d  &  30.7 d  \cr
PSR 1257+12 b, c &  66.5  &  1.48  &  15.2 min  &  22.3 min  \cr
Earth/Jupiter    &  365   &  11.9  &  1.42 min  &  24.1 s  \cr
\hline
\end{tabular}
\end{table}

\begin{figure}
\includegraphics[width=\fsize]{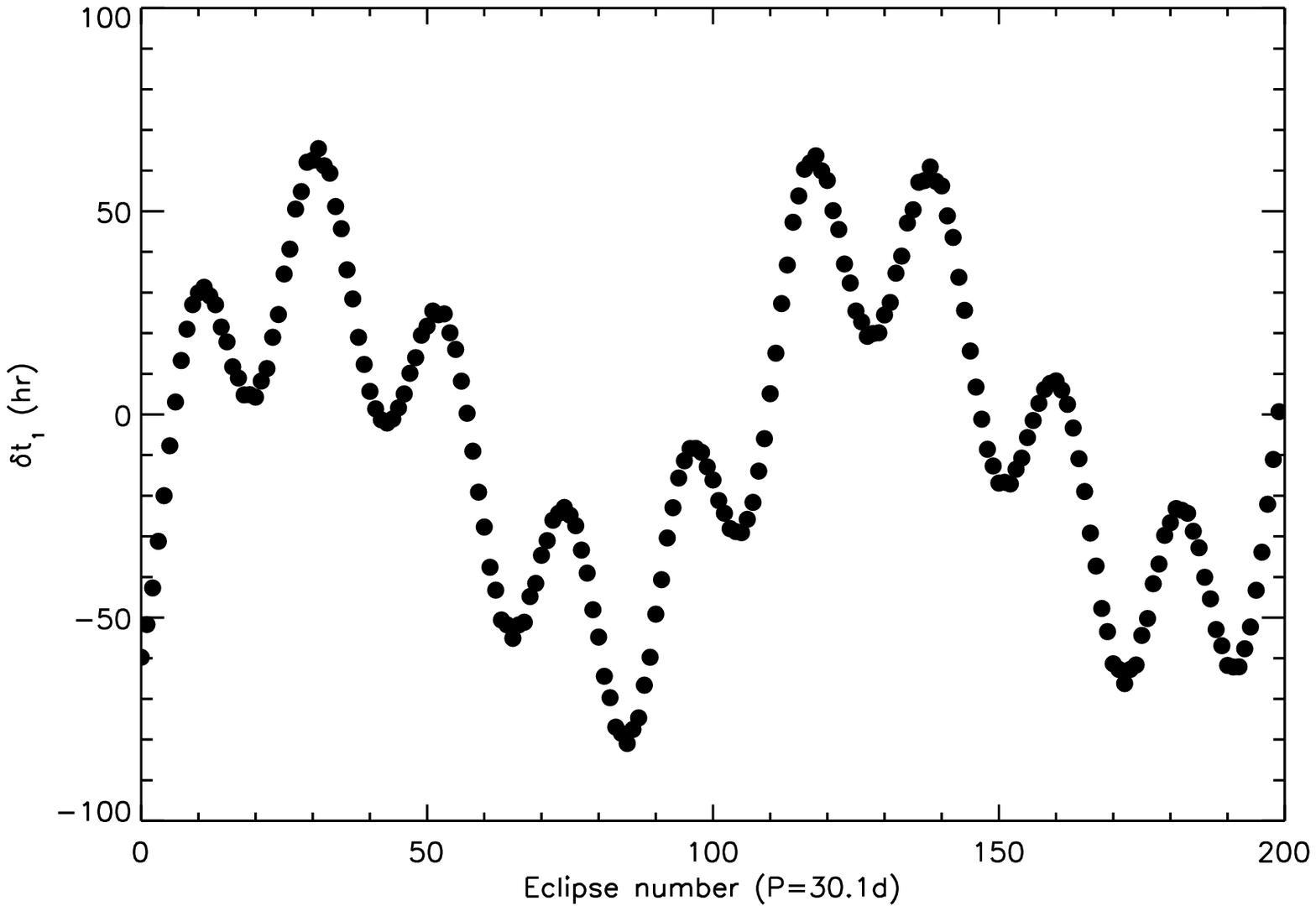}
\includegraphics[width=\fsize]{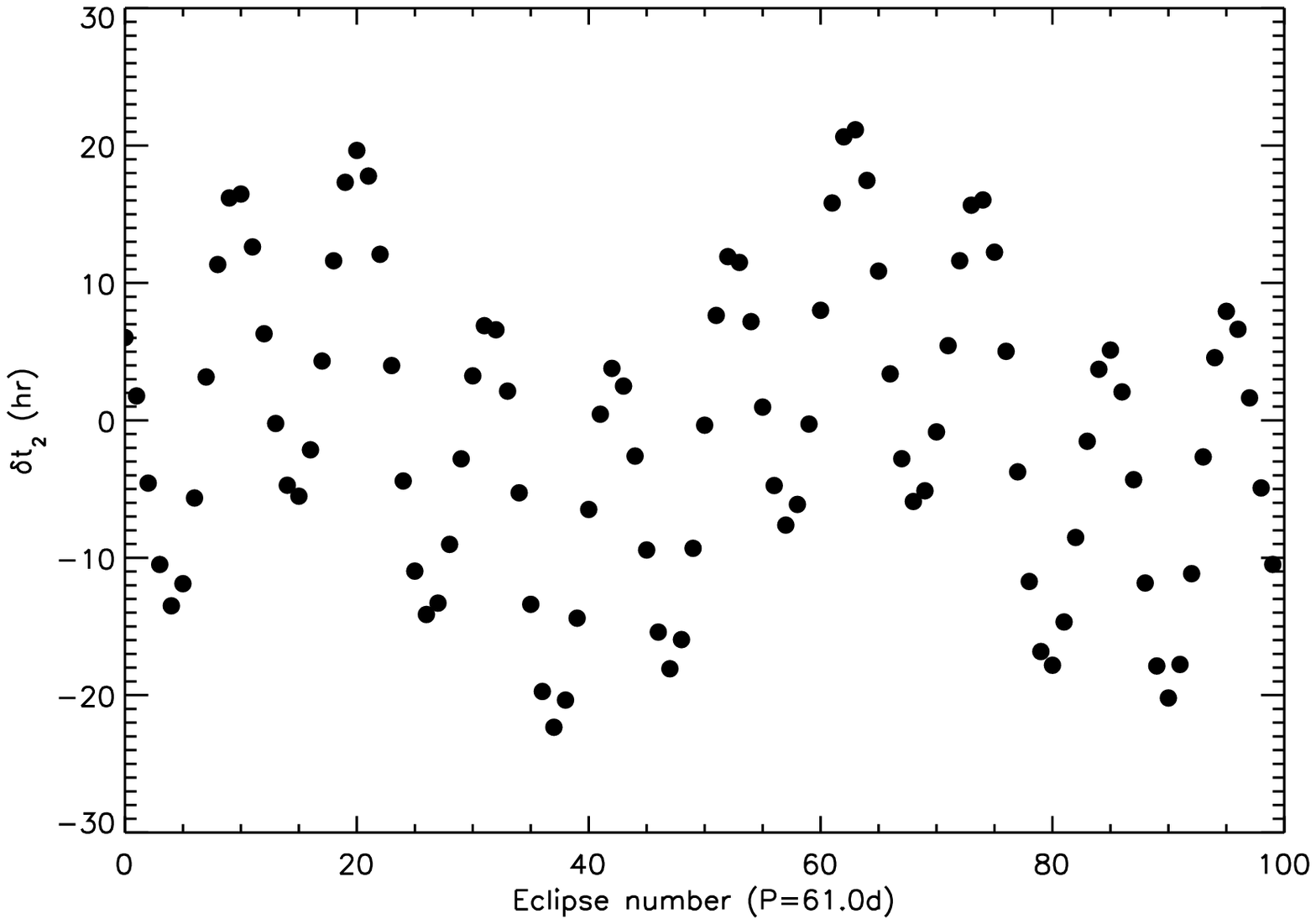}
\caption{(a) Transit timing variations for Gliese 876 B and
(b) Gliese 876 C.  The vertical axis is in units of hours, while
the horizontal axis is in units of the period of the transiting planet
(given in parentheses for reference).}
\label{fig6}
\end{figure}

The extrasolar planetary system Gliese 876 contains two Jupiter-mass
planets on modestly eccentric orbits which are near the 2:1 mean-motion resonance, 
$P_1=30.1$d and $P_2=61.0$d \citep{mar01}.  Due to the small size of the
M4 host star, the inner planet has a 1.5 per cent probability of transiting
for an observer at arbitrary inclination.  The orbital motion involves
both mean-motion resonance as well as a secular resonance in which the
planets librate about their apsidal alignment.  The apsidal alignment is
in turn precessing at a rate of $-41^{\circ}$ per year \citep{lau04,nau02,
riv01, lau01}.  Figure \ref{fig6} shows the predicted timing variations
if this system were seen edge-on and if the planets are coplanar using
the orbital elements from \cite{lau04}.

The two most prominent periodicities in Figure \ref{fig6} are those associated 
with the 
2:1 libration, with a period of roughly 600 days \citep[20 orbits of the inner 
planet,][]{lau01},
and the long term precession of the apsidal angle with a period of about 3200
days (110 orbits of the inner planet, corresponding to $-41^{\circ}$ yr$^{-1}$).
Evaluating equation (\ref{maxprecession}) gives a peak
amplitude of 1.4 days for the inner planet and 18 hours for the outer
planet which both compare well with the numerical 
results given that the eccentricities are not constant.

The extrasolar planetary system 55 Cancri contains a set of planets, $b$ and $c$,
near the $3$:$1$ resonance having 15 and 45 day periods.
There is some evidence for another planet, $d$, in an extremely long orbit,
and recently a fourth low mass planet, $e$, was found with a 2.8 day period
\citep{mca04}.  The planets $e$, $b$, and $c$ have transit probabilities of 
12, 4, and 2 per cent, respectively, for an observer at arbitrary inclination.  
The orbit of planet $b$ is approximately circular 
while planet $c$ is somewhat eccentric \citep{mar02}.  Table
1 gives the amplitude of the variations for the planets.
We have ignored planet $e$; however, it is at
a large enough semi-major axis to produce a $\sim$ 22 second variation 
due to light-travel time as the barycenter of the inner binary orbits
the barycenter of the triple system were the inner planets transiting.

The double planet system Upsilon Andromedae has a semi-major axis ratio 
of 14 which is not in a mean-motion resonance \citep{but99,mar01}.  The 
inner planet has a short period of 4.6 days, and thus a significant
probability of transiting of about 12 per cent, but has variations which are 
too small to currently be detected from the ground or space.  The outer 
planet has much larger transit timing variations due to its smaller 
velocity, but a much smaller probability of transiting.

The planetary system HD 37124 has two planets with a period ratio of
$\sim 10$ and a period of the inner planet of 241 days \citep{vog00}.  
The outer planet is highly eccentric, $e_2 =0.69$, and so its periapse 
passage produces a large and rapid change in the transit timing of the 
inner planet.
If this system were transiting, the variations would 
be large enough to be detected from the ground.  HD 82943 is in a 2:1
resonance giving variations of order the periods of the planets.
The pulsar planets are near a 3:2 resonance, which would cause large
transit timing variations were they seen to transit the pulsar progenitor
star.  Finally, alien civilizations observing transits of the Sun by 
Jupiter would have to have $\sim$ 10 second precision to detect the effect
of the Earth.

\section{Fourier Search for Semimajor Axis Ratio\label{fouriersearch}}

Since the TTV signal is typically periodic, it may be possible to use a discrete Fourier transform (FT) of the timing deviations to identify several of the orbital elements.  An analysis conducted with a Fourier representation of the data may even be more appropriate for finding the semimajor axis ratio of the system than analyzing the data in the time basis because the orbit of the periodic nature of the systems involved.  Other orbital elements may also be determined with the Fourier representation, though such a development is left for elsewhere.

Here, I present a method to determine the semimajor axis ratio of the planetary system using a Fourier transform of the timing deviations.  The value of this technique is that it could reduce the parameter space by one parameter entirely or, at least, provide several individual values about which a more complete search can be conducted.  This technique may also give an independent confirmation of an estimate for the semimajor axis ratio parameter that is obtained by another means.

\subsection{Fourier Representation}

\begin{figure}
\includegraphics[width=\fsize]{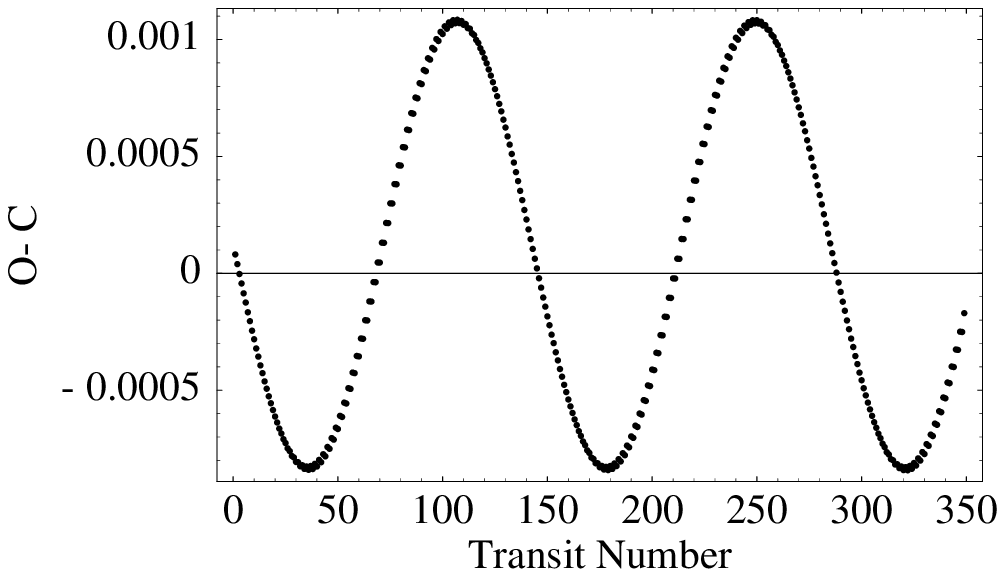}
\includegraphics[width=\fsize]{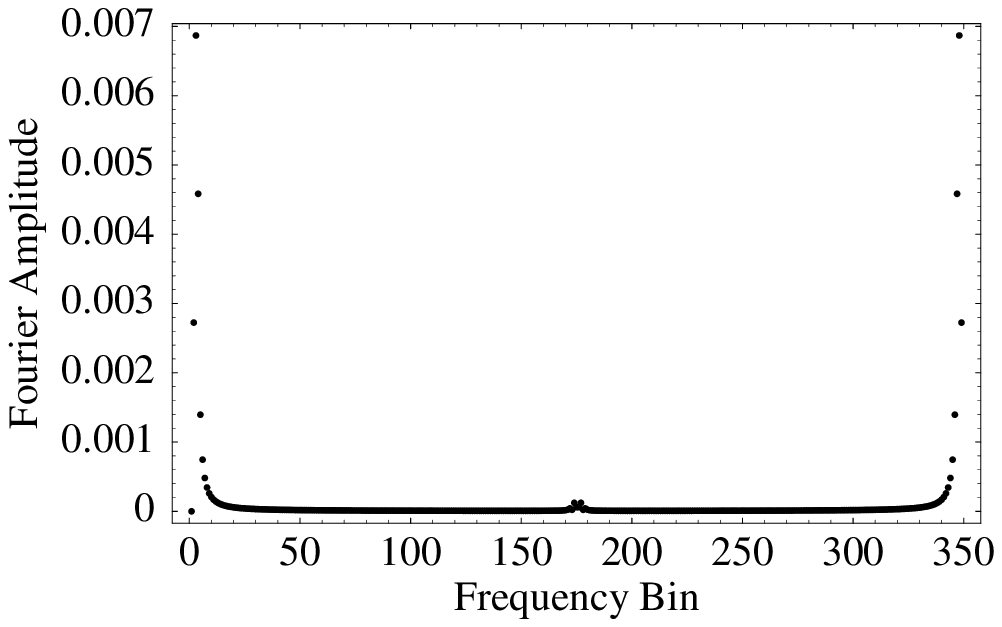}
\caption{Timing residuals (upper panel) with Fourier transform of those residuals (lower panel) for a planetary system that is near a mean-motion resonance.}
\label{fourres}
\end{figure}

\begin{figure}
\includegraphics[width=\fsize]{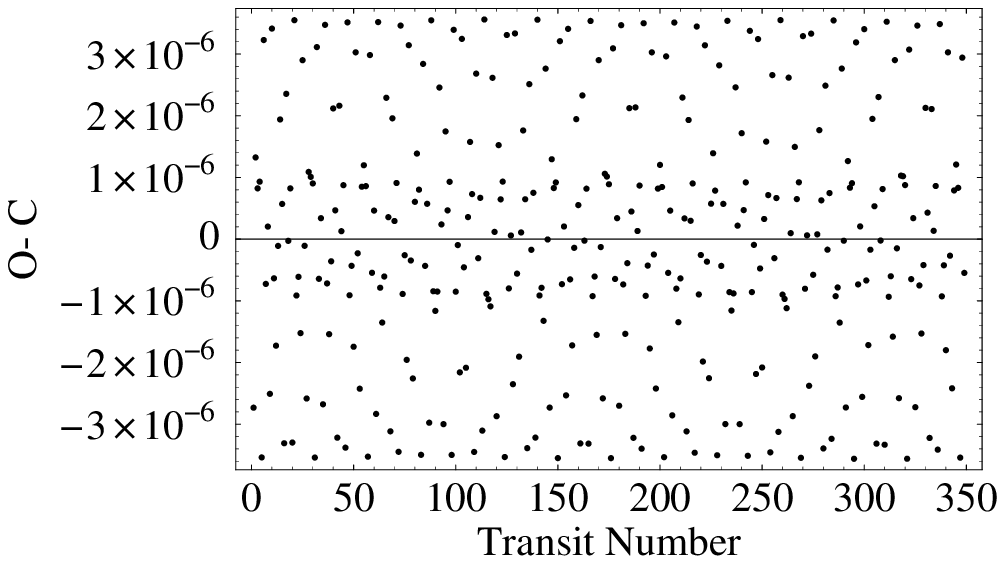}
\includegraphics[width=\fsize]{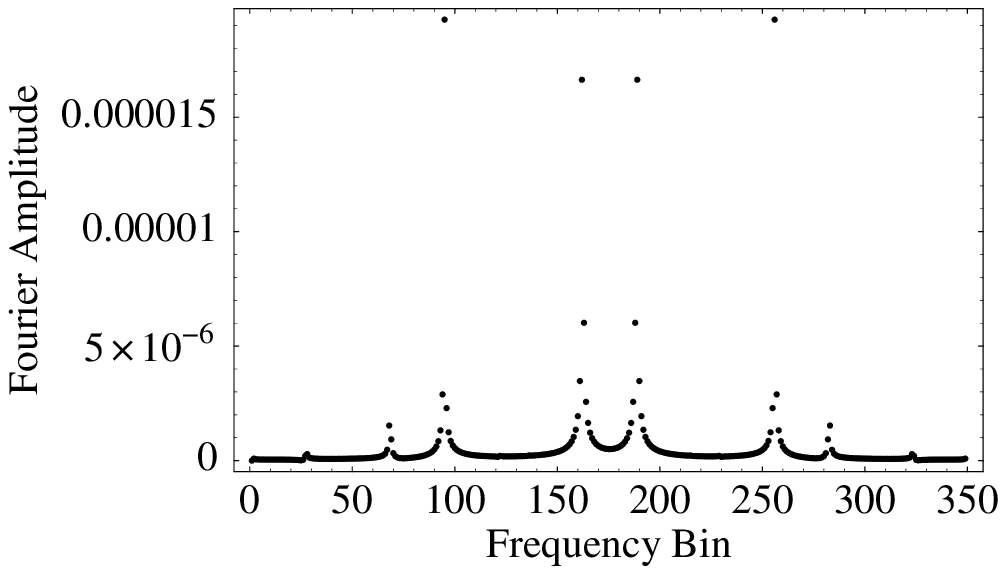}
\caption{Timing residuals (upper panel) with Fourier transform of those residuals (lower panel) for a planetary system that is not near a mean-motion resonance.}
\label{fournonres}
\end{figure}

Consider figures \ref{fourres} and \ref{fournonres}.  Figure \ref{fourres} shows the timing residuals of the transits of a planet where the perturber is near a mean-motion orbital resonance (the 2:1 resonance in this case).  The second image in this picture shows the FT of those residuals.  The symmetry in the Fourier transform is due to the fact that the timing information is encapsulated in both the amplitude and the phase of the Fourier components so that only the Fourier components up to the Nyquist frequency (half-way across the graph) contain unique information.  For this case the Fourier component that corresponds to the 2:1 resonant forcing term has the largest amplitude.  By comparison, figure \ref{fournonres} is for a system that is not near an orbital resonance and has several large peaks in the amplitudes of the Fourier components.

In order to determine whether or not the relative spacing between the peaks in the FT was interesting, I generated a set of planetary systems where both planets had zero eccentricity, equal mass, equal time of pericenter passage, and with various, equally spaced semimajor axis ratios with the perturber being exterior to the transiting planet.  For each system I tabulated the variations in transit time and then took the FT of the timing residuals.  I then calculated the time intervals that correspond to the largest peaks (up to five) of the FT.

\begin{figure}
\includegraphics[width=\fsize]{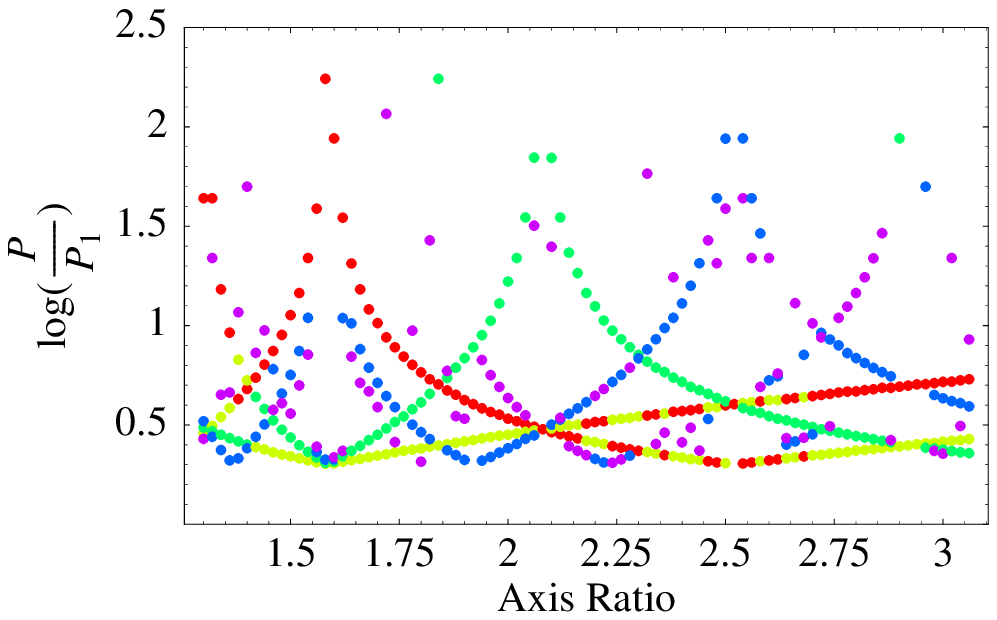}
\caption{This picture shows the periods that correspond to the five largest peaks in the Fourier transform of the timing residuals for a transiting planet where both planets are on initially circular orbits as a function of the ratio of the semimajor axes of the planets.  The period that corresponds to the highest peak is in red, the second highest is in yellow, then green, blue, and purple.\label{thing1}}
\end{figure}

From these results I generated a scatter plot of the values of the five time intervals as a function of the semimajor axis ratio and obtained figure \ref{thing1}.  Each of the visible peaks corresponds to a particular mean-motion resonance---the shape of which is given later.  One of the remaining structures is a diagonal line that corresponds to the period of the perturbing planet in units of the period of the transiting planet.  This diagonal line is particularly interesting because it is invertable and can, therefore, be used to uniquely identify the period of the perturbing planet.

\begin{figure}
\includegraphics[width=\fsize]{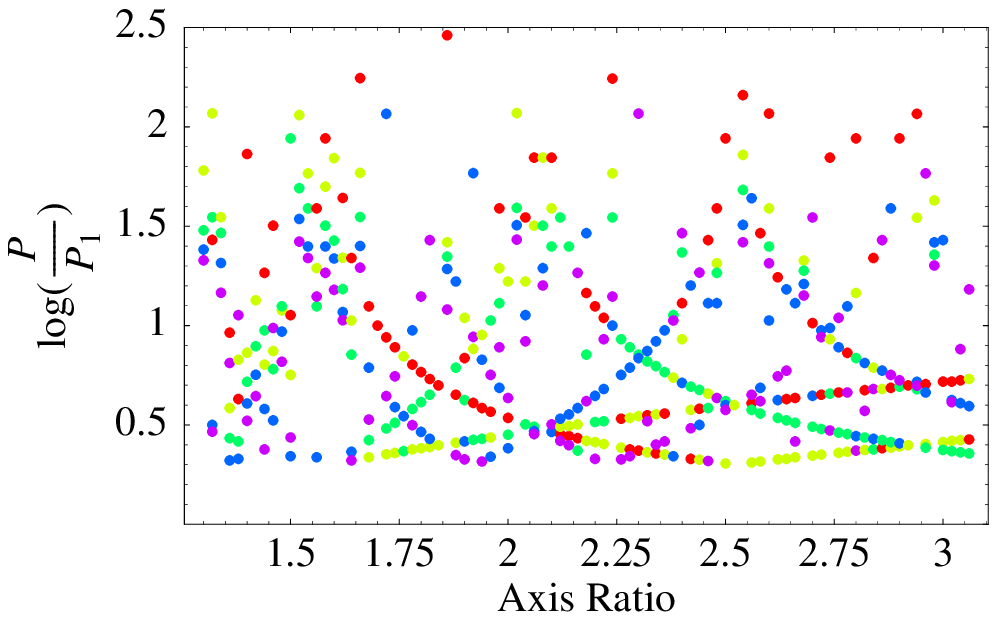}
\caption{Similar to Figure \ref{thing1}, this shows the periods that correspond to the five largest peaks in the Fourier transform of the timing residuals as a function of the semimajor axis ratio of the two planets.  Here the orbital elements of the systems at each axis ratio are randomly selected.  Red corresponds to the highest peak, then yellow, green, blue, and purple.\label{thing2}}
\end{figure}

The overall structure of this graph is virtually independent of the orbital elements of the system.  For example, another set of systems, with identical values of the semimajor axis ratio but with the remaining orbital elements being randomly generated results in Figure \ref{thing2}.  The primary difference between Figures \ref{thing1} and \ref{thing2} is that the relative heights of the various peaks in the FT; the peaks themselves remain in the same locations.

\begin{figure}
\includegraphics[width=\fsize]{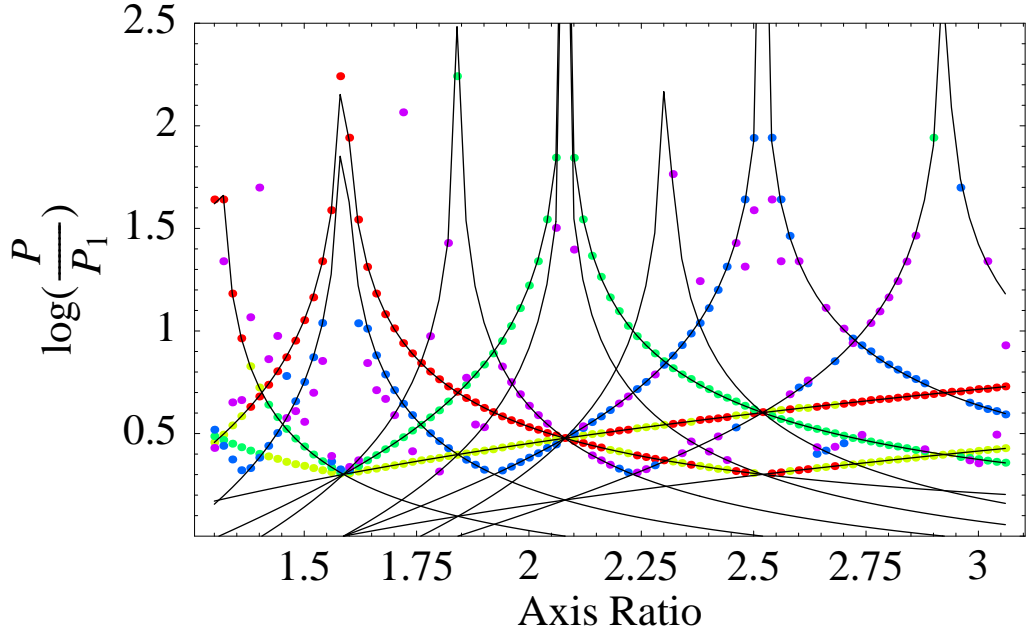}
\caption{This figure has the plot from Figure \ref{thing1} with several curves of equation \ref{lines} superposed.  Also shown are the two diagonal lines that correspond to the period of the perturbing planet (upper diagonal line) and half the period of the perturbing planet (lower diagonal line).\label{thing3}}
\end{figure}

The shape of the peaks that are visible in these two plots can be derived from the corresponding mean motion resonance.  For the $i$:$j$ resonance, where $i$ and $j$ are integers and where $iP_{\text{trans}}=jP_{\text{pert}}$ corresponds to exact resonance, the period that describes the shape of the structures in Figures \ref{thing1} and \ref{thing2} is given by
\begin{equation}\label{lines}
\frac{P_{\text{peak}}}{P_{\text{trans}}} = \frac{P_{\text{pert}}}{iP_{\text{trans}}-jP_{\text{pert}}}.
\end{equation}
This shows that the period of the Fourier component that appears on a given peak approaches infinity as the system approaches the resonance and falls away as the distance from that resonance increases.  Figure \ref{thing3} superposes several curves of equation (\ref{lines}), that correspond to different resonances, onto Figure \ref{thing1}.

\begin{figure}
\includegraphics[width=\fsize]{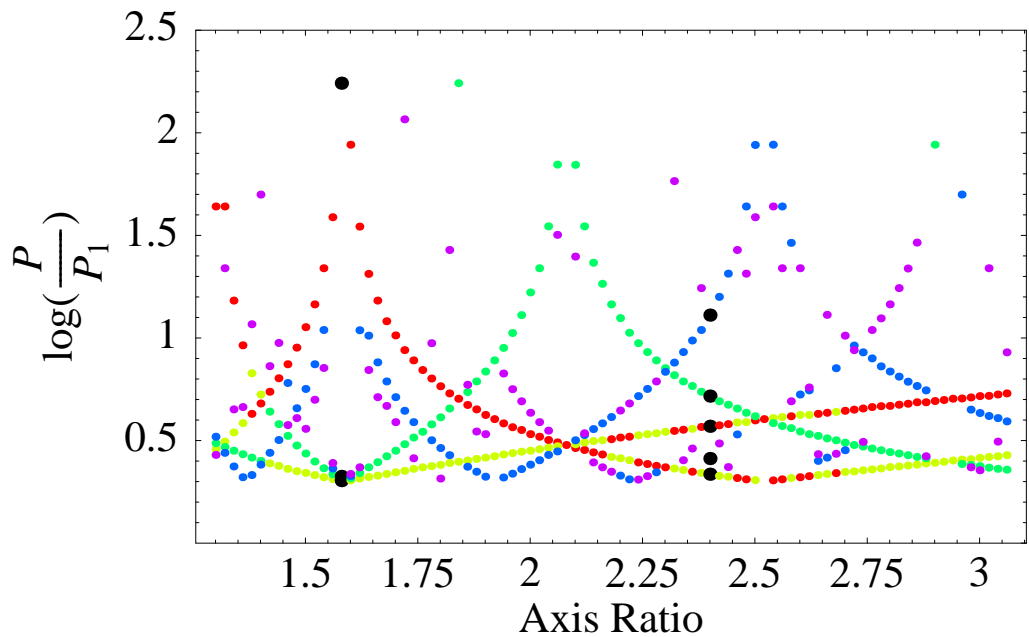}
\caption{This figure has the plot from Figure \ref{thing1} with the periods that correspond to the highest peaks in the Fourier transform---shown in black---of the resonant system whose transit times are shown in Figure \ref{fourres} (on the left) and the nonresonant system whose transit times are shown in Figure \ref{fournonres}.\label{thing4}}
\end{figure}

The value of this development is that the Fourier information can be used as a fingerprint to identify the period of the perturbing planet directly from the transform of the timing residuals.  Starting with the FT of a given set of timing residuals we plot the periods that are associated with the peaks in the amplitude of the FT.  We then assume that one of the peaks is precisely the period of the perturbing planet and compare the remaining peaks with the set of periods that result from equation (\ref{lines}).  For example, in Figure \ref{thing4} I show the periods that correspond to the peaks in the FT of the systems shown in Figures \ref{fourres} and \ref{fournonres} superposed upon Figure \ref{thing1}.

\subsection{Implementation}

Several tasks remain to be done before this technique to determine the semimajor axis ratio of the planetary system can be implemented in a robust and automated manner.  The largest impedement is that I have not found an appropriate statistical test to estimate the proper fit of the Fourier peaks to equation (\ref{lines}).  I initially tried several systems ``by eye'' and found that I could correctly identify approximately 90\% of the systems---though no noise had been added to the transit timings.  Most of the remaining systems had several possibilities that looked as though they fit equally well.

The difficulty lies in the fact that there are infinitely many resonances against which the data could be tested; but they are not equally important.  One could probably find a perfect fit to the Fourier peaks if he were allowed to include the effects of resonances such as the 6:3 resonance---which corresponds to the 2:1 resonance but still provides an independent curve from equation (\ref{lines}) (this can be seen in Figure \ref{thing1})---or the 200:101 resonance which is also near the 2:1 resonance but which is intuitively less important.  There is no clear method to determine which resonances should be weighted more heavily than others.  One potentially fruitful method would be to weight each higher order resonance (where the order is defined by $|i-j|$) by the maximum, average, or product of the eccentricities of the planets.  Thus, for low-eccentricity systems the importance of the high-order resonances would be severely limited; while for highly eccentric systems they would be more important.

Another challenge is that there are not always multiple peaks that can be identified in the FT---see the Figures \ref{fourres} and \ref{fournonres}.  Any noise that is added to the data will tend to wash out one or more of the peaks regardless of its importance to determining the semimajor axis ratio parameter.  For example, in the resonant system of Figure \ref{fourres}, the second largest peak corresponds to the (unique) period of the perturbing planet.  This peak is significantly smaller than the peak that corresponds to the 2:1 resonance; but without it there is no way to identify that the resonance is indeed 2:1---any resonance would be a candidate.  I have yet to determine an appropriate way to handle this issue and the one discussed in the preceeding paragraph.  Until such time as these problems have been solved the value of this Fourier technique is uncertain.  It does, however, illustrate the potential that the technique posseses.

\chapter{Applications and Caveats\label{apps}}



\section{Detection of terrestrial planets}\label{tpdet}

The possibility of detecting terrestrial planets using the transit
timing technique clearly depends strongly on (1) the period
of the transiting planet; (2) the proximity to resonance of
the two planets; (3) the eccentricities of the planets.  
The detectability of such planets also depends on
the measurement error, the intrinsic noise due to stellar variability,
gaps in the observations, etc.  One requirement for 
the case of an external perturbing planet is that observations should be 
made over a time longer than the period of the timing variations, which
can be longer than the period of the perturbing planet.
Ignoring these complications, a rough estimate of detectability can be 
obtained from comparing the standard deviation of the transit timing with 
the measurement error.

It is worthwhile to provide a numerical example for the case of a hot Jupiter 
with a 3 day period that is perturbed by a lighter, exterior planet on a 
circular orbit in exact 2:1 resonance.  The timing deviation amplitude is of 
order the period (3 days) times the mass ratio (300) or about 3 minutes
(equation \ref{reemeqn}):
\begin{equation}
\delta t=3\left({m_{pert} \over m_\oplus}\right) {\rm min}.
\end{equation}
These variations accumulate over a time-scale of order the period (3 days) times
the planet to star mass ratio to the power of $2/3$, which for a
transiting planet of order a Jupiter mass is about 5 months (equation \ref{plib}):
\begin{equation}
t_{cycle}= 150\left(m_{trans} \over m_J \right)^{-2/3} {\rm days}.
\end{equation}
Such a large signal should easily be detectable from the ground.
With relative photometric precision of $10^{-5}$ from space or from
future ground-based experiments, less massive objects or 
objects further away from resonance could be detected.  The observations could be
scheduled in advance and require a modest amount of observing time with the
possible payoff of being able to detect a terrestrial-sized planet.

\subsection{Comparison to other terrestrial planet search techniques}

To attempt a comparison with other planet detection techniques, we have
estimated the mass of a planet which may be detected at an amplitude
of 10 times the noise for a given technique.  We compare three
techniques for measuring the mass of planets: (1) radial velocity
variations of the star; (2) astrometric measurements; (3) transit
timing variations (TTV).  We assume that radial velocity measurements have
a limit of 0.5 m/s RMS uncertainty, which is about the 
highest precision that has been
achieved from the ground, and may be at the limit imposed by stellar
variability \citep{but04}.  We assume that astrometric measurements
have a precision of 1 $\mu$arcsecond which is the precision which
is projected to be achieved by the {\it Space Interferometry Mission}
\citep{for03,soz03}.  Finally, we assume that the transit timing
can be measured to a precision of 10 seconds, which is 
the highest precision of transit timing measurements of HD209458 
\citep{bro01}. 

\begin{figure}
\includegraphics[width=\fsize]{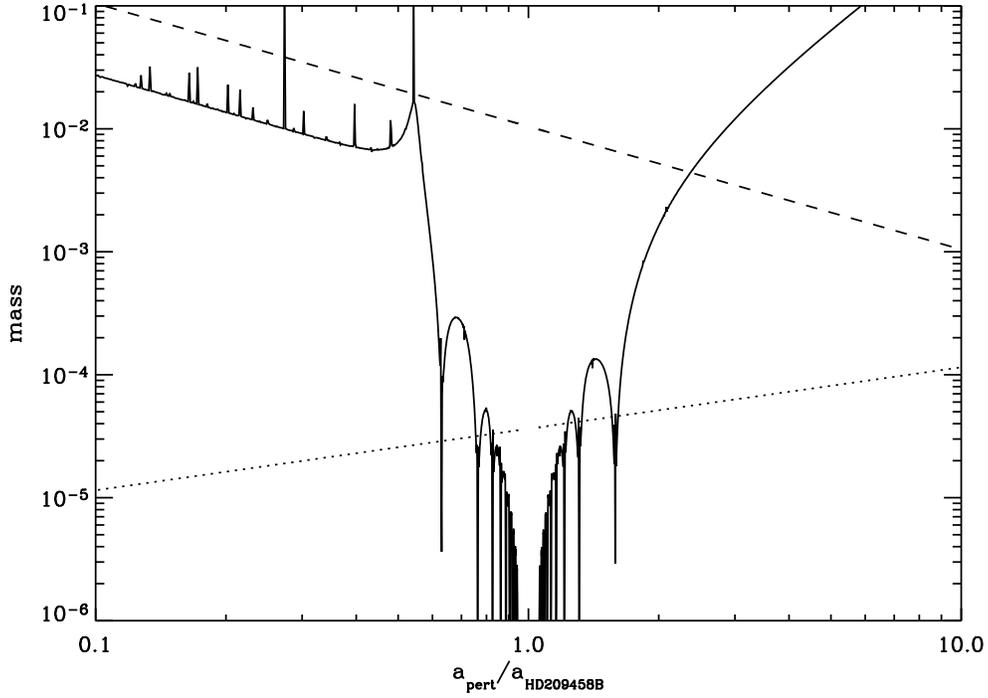}
\caption{Greatest lower bound on the detectable mass of a secondary planet for 
various planet detection techniques for the HD209458 system.  
The vertical axis is the perturbing
planet's mass in units of ${\rm M_\odot}$.   The horizontal axis is
the period ratio of the planets.  The solid line is for the transit timing
technique, the dashed line is astrometric, and the dotted line is the radial
velocity technique.}
\label{fig7}
\end{figure}

We concentrate on HD209458 since it is the best studied transiting 
planet.  This system is at a distance of 46 pc and has a period of 3.5 days.
Figure \ref{fig7} shows a comparison of the sensitivity
of these three techniques for a signal to noise ratio of 10.  
The solid curve is computed for
$m_{trans}=6.7\times 10^{-4} {\rm M_\odot}$ and $m_{pert}=10^{-7} {\rm M_\odot}$
and both planets on circular orbits.  When not trapped in resonance, the 
amplitude of the timing variations scales as $m_{pert}/m_0$, so we scale
the results to the mass of the perturber to compute where the timing
variations are 100 seconds -- this determines the sensitivity.  
The TTV technique is more sensitive than
the astrometric technique at semi-major axis ratios smaller than
about 2.  Note that in Figure \ref{fig7} the TTV 
and astrometric techniques have the same slope at small $P_{pert}/P_{trans}$.
This is because the transit timing technique is measuring 
the reflex motion of the host star due to the inner planet, which is
also being measured by astrometry and that the solid curve is an {\it upper limit} to the minimum mass detectable in HD209458 since a non-zero eccentricity will lead to larger timing variations (Figure \ref{fig5}) and thus a smaller detectable mass.

We see from \ref{fig7} that, off resonance, radial velocity measurements are the technique
of choice for this system, while on resonance the TTV is sensitive
to much smaller planet masses.  The improvement in sensitivity with the TTV technique over the radial velocity technique for a typical $j$:$j+1$ resonance is estimated as follows.  The radial velocity that a planet induces in its host star scales as $v \sim (a/P) (m_{\rm rv}/m_0)$ where $a$ is the semimajor axis, $P$ is the period, and $m_{\rm rv}$ and $m_0$ are the planet and star masses respectively.  By replacing the velocity of the star with the radial velocity measurement precision ($v \rightarrow \sigma_v$) and solving for the mass of the planet, we see that the radial velocity technique is able to detect planets with masses of order $m_{\rm rv} \sim \sigma_v m_0 P/a$.

Similarly, the typical timing deviations for a resonant system is given by equation \ref{reemeqn} with is approximately $\delta t \sim (P/10) (m_{\rm ttv}/m_p)$ where $P$ is the period, $m_p$ is the transiting planet mass, and $m_{\rm ttv}$ is the perturbing planet mass which I assume is much smaller than $m_p$.  The factor of 10 comes from the denominator of equation \ref{reemeqn}.  By replacing the transit timing deviation by the timing measurement uncertainty $\delta t \rightarrow \sigma_t$ and solving for the mass of the perturber we obtain the detection limit for the TTV technique $m_{\rm ttv} \sim 10 \sigma_t m_p/P$.  By dividing the mass limit for TTV with the mass limit for RV we find
\begin{equation}
\frac{m_{\rm ttv}}{m_{\rm rv}} \sim \frac{10\sigma_t a}{\sigma_v P^2}\left(\frac{m_p}{m_0}\right) \sim ({\rm few}) 10 \frac{m_p}{m_0}
\end{equation}
for typical values of the different observational and transiting hot jupiter parameters.  The large improvement (an order-of-magnitude or more) stems from the fact that the resonant TTV signal scales as the mass ratio of the planets because of conservation of the energy of the planetary orbits while the radial velocity scales with the planet to star mass ratio due to conservation of the barycenter motion.

\section{Breaking the mass-radius degeneracy}\label{mrdeg}

In the case that two planets are discovered to transit their host star,
measurement of the transit timing variations can break the degeneracy
between mass and radius needed to derive the physical parameters of the
planetary system.  This has been discussed by \cite{sea03a} who use 
a theoretical stellar mass-radius relation to break this degeneracy.  We 
provide a simplified treatment here to illustrate the nature of the 
degeneracy and how it can be broken with observations of transit timing
variations.

Consider a planetary system with two transiting planets on circular orbits
which are coplanar, exactly edge-on, and have measured radial velocity 
amplitudes.  We'll assume that the star is uniform in surface brightness
and that $m_1, m_2 \ll m_0$.  We'll also assume that the unperturbed periods 
$P_1, P_2$ can be measured from the duration between transits.
Then there are eight physical parameters of interest which
describe the system: $m_0, m_1, m_2, R_0, R_1, R_2, a_1,$ and $a_2$ where $R_i$ 
are the radii of the star and planets.  Without
measuring the transit timing variations, there are a total of ten parameters
which can be measured:  $K_1, K_2, t_{T1}, t_{T2}, t_{g1}, t_{g2}, \Delta F_1,
\Delta F_2, P_1,$ and $P_2$, where $t_{Tj}$ labels the duration of transit
from mid-ingress to mid-egress, $t_{gj}$ labels the duration
of ingress or egress for planet $j$, $K_j$ are the velocity amplitudes of
the two planets, and $\Delta F_j$ are the relative depths of the transits in
units of the uneclipsed brightness of the star (for planet $j$).  Although 
there are more
constraining parameters than model parameters, there is a degeneracy since
some of the observables are redundant.  All of the system
parameters can be expressed in terms of observables and the ratio of
the mass to radius of the star, $m_0/R_0$,
\begin{eqnarray}
{R_j \over R_0} &=& \Delta F_j^{1/2}\cr
{m_j \over m_0} &=& K_j \left({P_j \over \pi t_{Tj} G}\right)^{1/2}  \left({m_0\over R_0}\right)^{-1/2} \cr
a_j &=& {P_j \over 2 \pi} \left({G \pi t_{Tj} \over P_j }\right)^{1/2} \left({m_0\over R_0}\right)^{1/2} \cr
{R_0 \over a_j} &=& {\pi t_{Tj} \over P_j}\cr
{R_j \over a_j} &=& {\pi t_{gj} \over P_j},
\end{eqnarray}
where $j=1,2$ labels each planet.
From this information alone one can constrain the density of the star 
\citep{sea03a}.  For the simplified case discussed here,
\begin{equation}
\rho_* = {3 P \over \pi^2 G t_T^2}
\end{equation}
for either planet \citep[this differs sligthly from the expression in][since we define
the transit duration from mid in/egress]{sea03a}.
If, in addition, one can measure the amplitude of the transit timing variations
of the outer planet, $\sigma_2$, then this determines the mass ratio.  For the
case that the star's motion dominates the transit timing,
\begin{equation}
{m_1\over m_0} = {2 \sqrt{2} \pi \sigma_2 \over P_2^{1/3} P_1^{2/3}}.
\end{equation}
For other cases, the transit timing amplitude can be computed numerically.
Then, from the above expression for $m_i/m_0$ one can find the
ratio of the mass to the radius of the star
\begin{equation}
{m_0 \over R_0} = {1 \over 8\pi^3 G} {P_1^{7/3} P_2^{2/3} K_1^2 \over t_{T1} \sigma_2^2}.
\end{equation}
Combined with the measurement of the density, this gives the absolute
mass and radius of the star.  This
procedure requires no assumptions about the mass-radius relation for the
host star, and in principle could be used to measure this relation.
If one can also measure transit timing variations for the inner planet,
then an extra constraint can be obtained
\begin{equation}
\sigma_1 = {P_1 m_2 \over m_0} f(\alpha),
\end{equation}
where $f(\alpha)$ is a function derived from averaging equation
\ref{sigcirc1}.  (Note that the phase of the orbits is needed
for this equation, which can be found from the velocity amplitude
curve).  This provides an extra constraint on the system,
and thus will be a check that this procedure is robust.

Clearly we have made some drastically simplifying assumptions which
are not true for any physical transit.  The inclination of
the orbits must be solved for, which can be done from the ratio
of the durations of the ingress and transit and the change in flux,
as demonstrated by \cite{sea03a}.
In addition, limb-darkening must be included, and can
be solved for with high signal-to-noise data as demonstrated by
\cite{bro01}.  Finally, the orbits are not generally
circular, so the parameters $e_j, \varpi_j, \Omega_j, \sigma_j$, which 
can be derived from the velocity amplitude measurements,
should be accounted for.
The general solution is rather complicated and would best be
accomplished numerically, but the degeneracy has a similar
nature to the circular case and can in principle be broken by the transit
timing variations.

\section{Effects we have ignored}

I now discuss several physical effects that I have ignored but which
ought to be kept in mind by observers measuring transit timing
variations.

\subsection{Light travel time} \label{lighttravel}

\cite{dee00} carried out a search for perturbing planets
in the eclipsing binary stellar system, CM Draconis, using the changes
in the times of the eclipse due to the light travel time to measure 
a tentative signal consistent with a Jupiter-mass planet at $\sim 1$ AU 
(their technique would in principle be sensitive to a planet on
an eccentric orbit as well, c.f equation \ref{eccentricouter}).
The ``R\o mer Effect'' due to 
the change in light travel time caused by the reflex motion of the 
inner binary is much smaller in planetary systems than in binary stars
since their masses and semi-major axes are small, having an amplitude
\begin{equation}
t_{star}= {a\over c} {m_p\over m_*} \approx 0.5 {\rm sec} \left({m_p \over M_J}\right) \left({a_2 \over 1 {\rm AU}}\right) 
\end{equation}
where $M_J$ is the mass of Jupiter and $a_2$ is the semi-major axis of
the perturbing planet.  This effect is present in the absence of 
deviations from a Keplerian orbit because the inner binary orbits about the center of mass.

There can also be changes in the timing of the transit as
the distance of the transiting planet from the star varies.
In this case, the time of transit is delayed by the light travel time
between the different locations where the planet intercepts the beam of light from
the star.  The amplitude of these variations is smaller than the $\sigma$ 
we have calculated by a factor of $\sim v_{trans}/c$,
where $v_{trans}$ is the velocity of the transiting planet.  So, only
very precise measurements will require taking into account light travel
time effects, which should be borne in mind in future experiments (of course
the light-travel time due to the motion of the observer in our solar system 
must be taken into account with current experiments).

\subsection{Inclination}

We have assumed that the planets are strictly coplanar and exactly edge-on.
The first assumption is based on the fact that the solar system is nearly
coplanar and the theoretical prejudice that planets forming out
of disks should be nearly coplanar.  Small non-coplanar effects will
change our results slightly \citep{mir02},  while large inclination effects 
would require a reworking of the theory.  Since some extrasolar planetary systems 
have been found with rather large eccentricities, it is entirely possible that 
non-coplanar systems will be found as well, a possibility we leave for future 
studies.

The assumption that the systems are edge-on is based on the fact that
a transit can occur only for systems that are nearly edge-on.  For small
inclinations our formulae will only change slightly, but may result in
interesting effects such as a change in the duration of a transit, or even
the disappearance of transits due to the motion of the star about
the barycenter of the system.  On a much longer time-scale (centuries), the 
precession of an eccentric orbit might cause the disappearance of transits
since the projected shape of the orbit on the sky can change.  This possibility
was mentioned by \cite{lau04} for GJ 876.

\subsection{Other sources of timing ``noise''}

Aside from the long term effects that have been ignored there are several 
sources of timing noise that must be included in the analysis of observations 
of transiting systems.  These sources of noise could come from
the planet or the host star.  If the planet has a moon or is a binary
planet then there is some wobble in its position causing a change in 
both the timing and duration of a transit \citep{sar99, bro01}.  A moon or ring 
system may transit before the planet causing a shallower transit to appear earlier 
or later than it would without the moon \citep{bro01,schult,bar04}.  A large
scale asymmetry of the planet's shape with respect to its center of mass
might cause a slightly earlier or later start to the ingress or end of egress.

Stellar variability could also make a significant contribution to the noise. 
Variations in the brightness of the star might affect the accuracy of the
measurement of the start of ingress and the end of egress, which are the
times that are critical to timing of a transit.
Stellar oscillations can cause variations in the surface
of the Sun of $\sim 100$ km in regions of size $10^{3-4}$ km, which corresponds 
to a one second variation for a planet moving at 100 km/s.

\subsection{Coverage Gaps}

With radial velocity measurements and prior transit lightcurves one can 
predict the epoch of future transits and identify appropriate times for 
photometric monitoring of the system of interest. Observational
limitations (e.g. bad weather, equipment failure, scheduling requirements)
will lead to transits being missed, which in turn will cause inaccuracies
in $\sigma$.  Since the signal is periodic, $\delta t (t)$ may be
straightforward to extract with a few missed transits; however, if
the outer planet is highly eccentric, then most of the change in
transit timing may occur for a few transits \citep[e.g. 
HD37124; a similar selection effect occurs in radial velocity searches
as discussed by][]{cum04}. In principle
this effect will average out over long observational intervals; however as
in this context ``long'' may mean several decades or more, it will be
important to evaluate the effect of coverage gaps on detections over a
time-scale of months-years.  We will return to this in detail in future 
work. 

We note in passing that the advent of the new astrometric all-sky surveys
such as Gaia \citep{per01} will provide photometry for $\sim 10^7$
stars $>$ 15 mag, with fewer coverage gaps than ground-based observations;  
we thus expect the detection method by transits alone (section \ref{tpdet}) to
really come into its own over the next two decades. Assuming $\sim$ 0.4
detections (three transits) per 10$^4$ stars \citep[c.f.][]{bro03}, we may
expect transit lightcurves of perhaps $\sim 1000$ exoplanetary systems
over the mission lifetime, greatly aiding the determination of $\sigma$ 
for these systems.  The NASA {\it Kepler} mission will also provide
uniform monitoring of about $10^{2-3}$ transiting gas giants \citep{borucki},
and if flown, the Microlensing Planet Finder (formerly {\it GEST}) will 
discover $10^{4-5}$ transiting gas giants with uniform coverage \citep{ben03}.

\section{Conclusions}

For an exoplanetary system where one or more planets transit the host star the 
timing and duration of the transits can be used to derive several physical 
characteristics of the system.  This technique breaks the degeneracy between 
the mass and radius of the objects in the system.  The inclination, absolute 
mass, and absolute radii of the star and planets can be found; in principle
this could be used to measure the mass-radius relation for stars that are
not in eclipsing binaries.

In addition, TTV can be used to infer the existence of previously undetected 
planets.  We have found that for variations which occur over several orbital 
periods the strongest signals occur when the perturbing planet is either in a 
mean-motion resonance with the transiting planet or if the transiting planet 
has a long period (which, unfortunately, makes a transit less probable).  
The resonant case is more interesting since the 
probability of a planet transiting decreases significantly as the semi-major 
axis becomes large.  Using the TTV scheme it is possible to detect earth-mass 
planets using current observational technologies for both ground based and 
space based observatories.  Observations for several transits of HD209458 could be 
gathered and studied over a relatively short time due to its
small period.  Once the existence of a second planet is established one can 
predict the times at which it would likely transit the host star.  Follow-up 
observations with {\it HST} or high-precision ground-based telescopes
at those times would increase the likelihood of 
detecting a transit of the second planet.  

If the second planet is terrestrial 
in nature, this transit timing method may be the only way currently to 
determine the mass of such planets in other star systems.  Astrometry is 
another possible technique but it may take a decade of technological development 
before the necessary sensitivity is achieved.  In addition, complementary techniques
are necessary to probe different parts of parameter space and to provide extra
confidence that the detected planets are real, given the likely low signal-to-noise
\citep{gou04}.  For the near future the TTV 
technique may be the most promising method of detecting earth-mass planets
around main sequence stars besides the Sun.

We exhort observers to (1) discover longer period transiting planets \citep{sea03b}
since the signal increases with transiting planet's period; (2) increase the
signal-to-noise of ground based differential photometry \citep{how03} for more
precise measurement of the transit times; and (3) examine their transit
data for the presence of perturbing planets \citep{bro01}.

The treatment of this problem has ignored many effects which we plan
to take into account in future works in which we will simulate realistic
lightcurves including noise and to fit the simulated data to derive 
the parameters of the perturbing planet, exploring degeneracies in the 
period ratio.  We will also derive the probability of detecting such systems 
taking into account various assumptions about the formation, evolution, and
stability of extrasolar planets.  At this stage we do not address the effects of general relativity since the timescale over which these effects would be manifest is significantly longer than the timescale over which existing and planned observations are to collect data.

\chapter{The TrES-1 System\label{tres1}}

In this chapter and the next I present two analyses of existing transit data.  The first data set is for the TrES-1 planetary system (GSC 02652-01324) and the second is for the HD209458 planetary system.  The transit data for the TrES-1 system were obtained using ground-based telescopes and were reported by David Charbonneau in \cite{char05} (hereafter C05).  The data for the HD209458 system were obtained with the Hubble Space Telescope.  Many of the results of this chapter are reported in \cite{steffenagol05}.

The timing data reported in C05 were derived from the 11 transits reported by \cite{alo04} with an additional transit that was observed at the IAC 80cm telescope after \cite{alo04} went to press.  One transit was excluded from their analysis because it constituted a 6-$\sigma$ departure from a constant period and because of anomalous features in the ingress and egress.  That point, if it is valid, is the most interesting point for our purposes because the TTV signal is defined by such deviations.  Consequently, I analyse two different sets of data from C05; the ``12-point'' set which includes this point, and the ``11-point'' set which does not.  This study was the first analysis of TTV as presented by \cite{agol05} and \cite{holm05}.  And, it may be the first search for planets around main-sequence stars that can probe masses smaller than the mass of the Earth.

\section{Search for Secondary Planets\label{search}}

I conducted a variety of searches for the best-fitting perturbing planet in the TrES-1 system.  These searches included different combinations of orbital elements for TrES-1b.  I found that any reduction in the overall \cs\ obtained by including the parameters $e$ and $\varpi$ for TrES-1b was offset by the loss of a degree of freedom.  Therefore, I report results from the search where the eccentricity of TrES-1b was fixed at zero.

For this analysis, I stepped through the semi-major axis ratio of the putative secondary planet and TrES-1b.  At each point I minimized over six parameters: the eccentricity, longitude of pericenter, time of pericenter passage, and mass of the secondary planet and the period and the initial longitude of TrES-1b.  The inclination and ascending node of the perturbing planet were identical to the values for TrES-1b.  I analysed the data for both interior and exterior perturbers.

\begin{figure}
\includegraphics[width=\fsize]{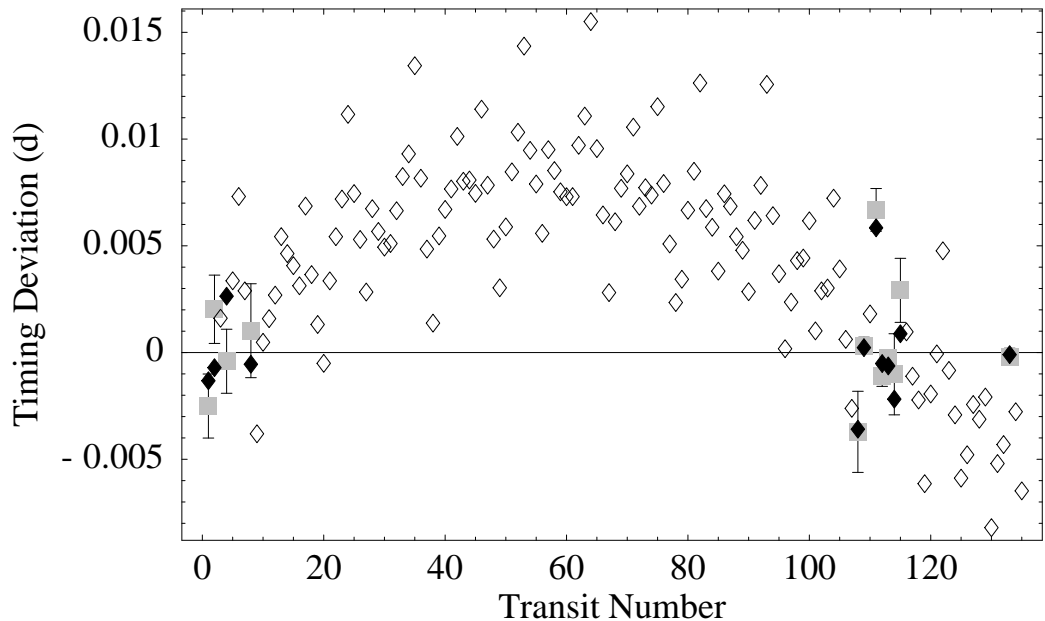}
\caption{A comparison of the TrES-1b transit timing residuals to those from the 12-point analysis for an interior perturber.  The diamonds are the timing residual at each transit, the filled diamonds denote transits where there is data, the squares with error bars are the data.  The 12-point solution has a reduced \cs\ of 2.8 on 6 degrees of freedom with ephemeris parameters $m = 0.111M_{\rm J}$, $P = 1.76d$, $e = 0.153$, $\varpi = 154.8^{\circ}$, $\tau = 2452847.4392d$.}
\label{timings1}
\end{figure}

This analysis did not produce any promising solution; though I present one interesting case for an interior perturber found from the 12-point analysis.  This solution is not near a low-order mean-motion resonance.  Indeed, the \cs\ was generally much higher near the $j$:$j+1$ and the $j$:1 resonances than in the regions between them.  Figure \ref{timings1} compares the timing residuals for this solution with the data.  The reduced \cs\ for this system is 2.8 on $N-6$ degrees of freedom (where $N=12$ is the number of data) compared with 6.3 for no perturber ($N-2$).

This solution, while it would have been detected from RV measurements, is interesting because the average size of the timing deviations is larger than that of the data---making the variations easier to detect.  However, I suspect that while this solution is numerically valid, it is merely an artifact of the gap between the two primary epochs of observation.  For this solution, and others like it, the simulated timing residuals consist of small, short-term variations superposed on a large-amplitude, long-period variation with a period that is a multiple of the difference between the two epochs.  Several candiate systems for both the 11-point and the 12-point analyses showed similar behavior.  Such results indicate one drawback of gaps in the coverage of transiting systems.  However, observing each transit of every transiting system may be neither feasable nor optimal for identifying perturbing planets with this technique given the limited resources that are appropriate for transit observations.  In the meantime, additional data for this system, taken at a time that is not commensurate with the existing gap in the observations would remove false solutions of this type in future studies.

From the above analysis, I conclude that there is not sufficient information in the data to uniquely and satisfactorally determine the characteristics of a secondary planet in the TrES-1 system.  This is in part because the number of model parameters is not much larger than the number of data and because the typical timing precision, $\sim 100s$, is not a sufficiently small fraction of the orbital period of the transiting planet (about $4\times 10^{-4}$) to distinguish between different solutions.  In addition, the gap in coverage appears to affect the minimization dramatically.  I believe that the coverage gap is primarily responsible for the observed fact that two nearby sets of orbital elements will have very different values of \cs ---a slight change in the long-term variation will cause the simulated transit times to miss the second epoch of observations.  Additional timing data, with precision comparable to the most precise of the given data, $\sim 30s$, and at an epoch that is not commensurate with the existing gap in coverage will allow for a more complete investigation of the system.

\section{Constraints on Secondary Planets}

That the results of the planet search were inconclusive is not particularly surprising since the transit timings show little variation from a constant period and the point that deviates the most is suspect.  If a satisfactory solution cannot be determined from these data, we can still use them to place constraints on the orbital elements of a two-planet system.  Of particular interest is a constraint on the mass that the secondary planet could have as a function of various orbital elements; similar to the constraint that resulted from analyzing the OGLE-1998-BUL-14 microlensing event~\citep{albrow00}.  As the mass of the hypothetical perturbing planet increases, the \cs\ of the model should grow significantly regardless of the values of the remaining orbital elements.  Therefore, I systematically studied a grid of values on the semi-major axis/eccentricity plane of the perturbing planet---assuming that the orbit of TrES-1b is initially circular.  At each point we identified the mass that a perturber needs in order to produce a large deviation from the data.

At each location in the $a$/$e$ plane we set the mass of the companion to be very small ($\ll M_{\oplus}$) and, for a random value for the longitude of pericenter and the time of pericenter passage, we calculated the \cs\ of the timing residuals.  We increased the mass of the secondary planet until the \cs\ grew by some fiducial amount.  At that point we minimized the \cs\ over the longitude of pericenter and the time of pericenter passage of the secondary planet and the initial longitude of TrES-1b.  Following the minimization the \cs\ typically fell below the fiducial amount and the mass of the perturber was again increased.  However, if after the minimization the \cs\ remained above the threshold, the orbital elements of the system were recorded, the \cs\ threshold increased, and the process repeated until either a maximum mass of the perturbing planet or a maximum \cs\ was achieved.  This procedure gives the minimum \cs\ as a function of perturbing planet mass for each point in the $a$/$e$ plane.

\begin{figure}
\includegraphics[width=\fsize]{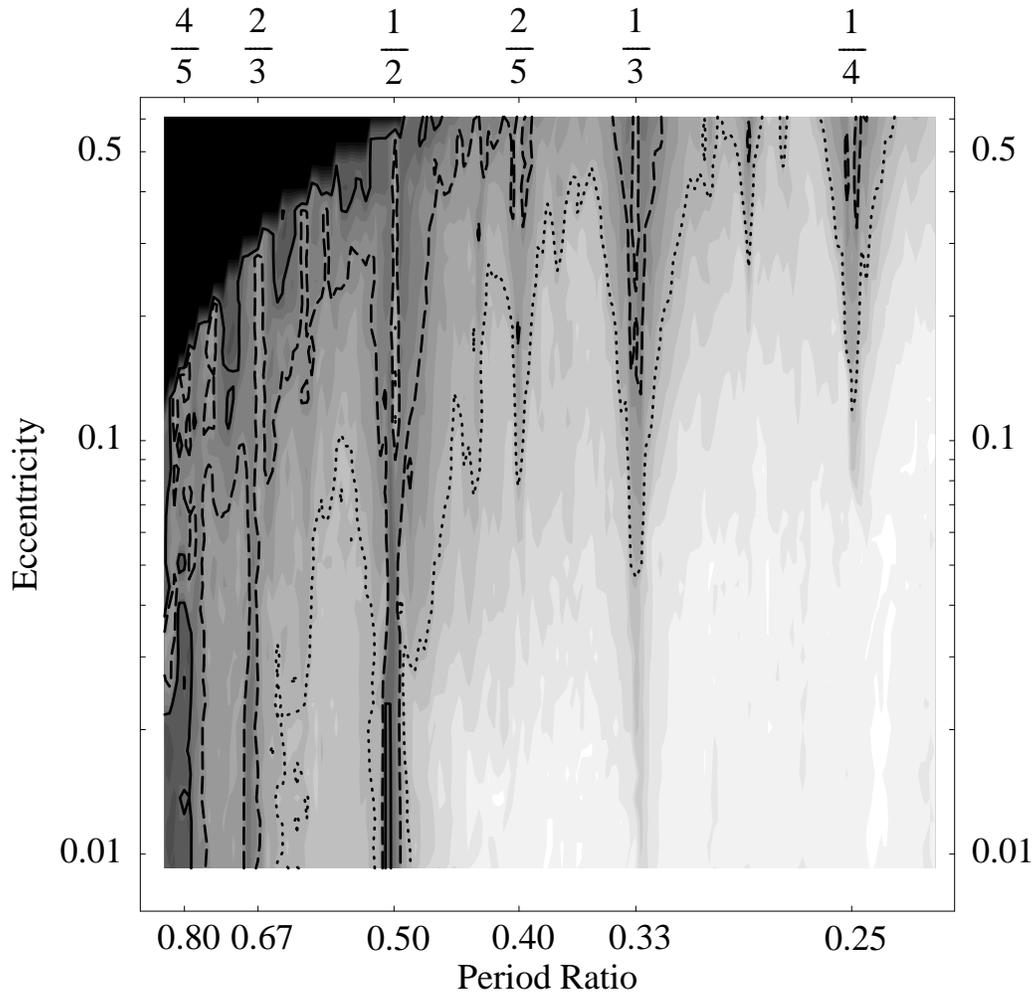}
\caption{The upper limit on the mass of an interior perturbing planet in the TrES-1 system from the 11-point analysis.  The regions bounded by the contours correspond to a companion mass that is less than 100 (dotted), 10 (dashed), and 1 (solid) earth masses.}
\label{3sigin}
\end{figure}

\begin{figure}
\includegraphics[width=\fsize]{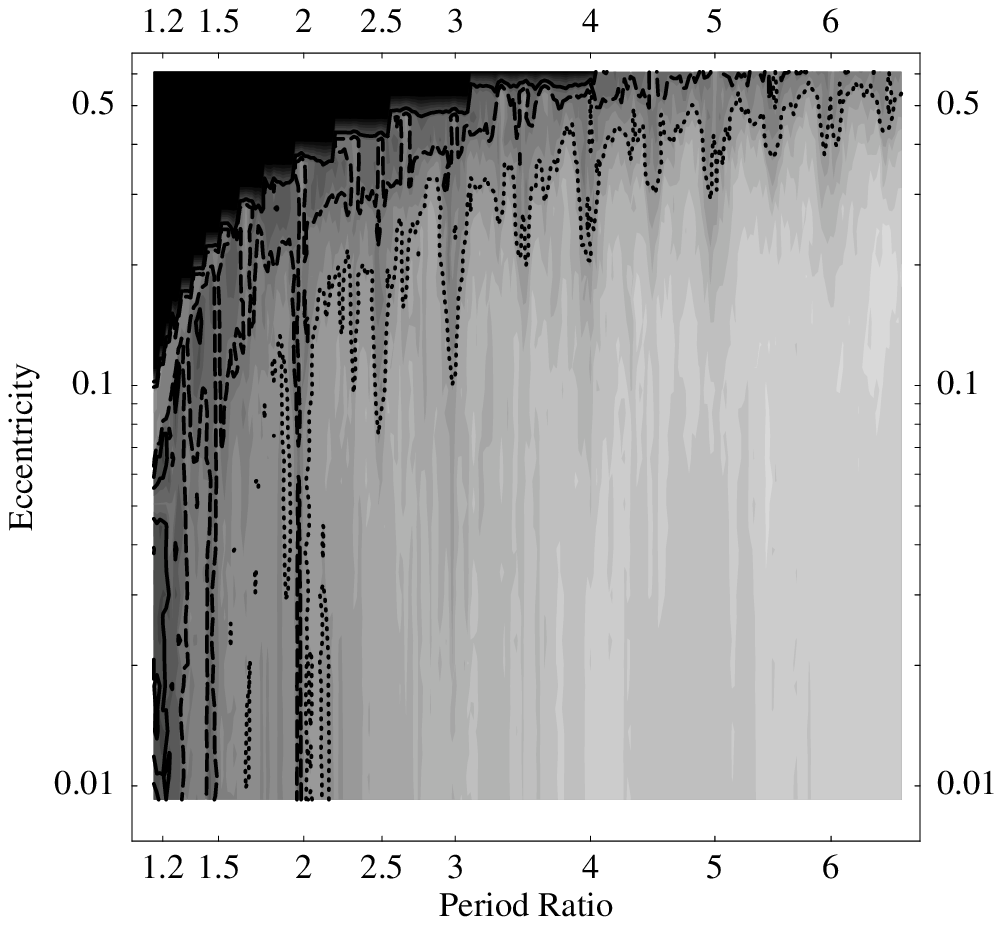}
\caption{The upper limit on the mass of an exterior perturbing planet in the TrES-1 system from the 11-point analysis.  The regions bounded by the contours correspond to a companion mass that is less than 100 (dotted), 10 (dashed), and 1 (solid) earth masses.}
\label{3sigout}
\end{figure}

\begin{figure}
\includegraphics[width=\fsize]{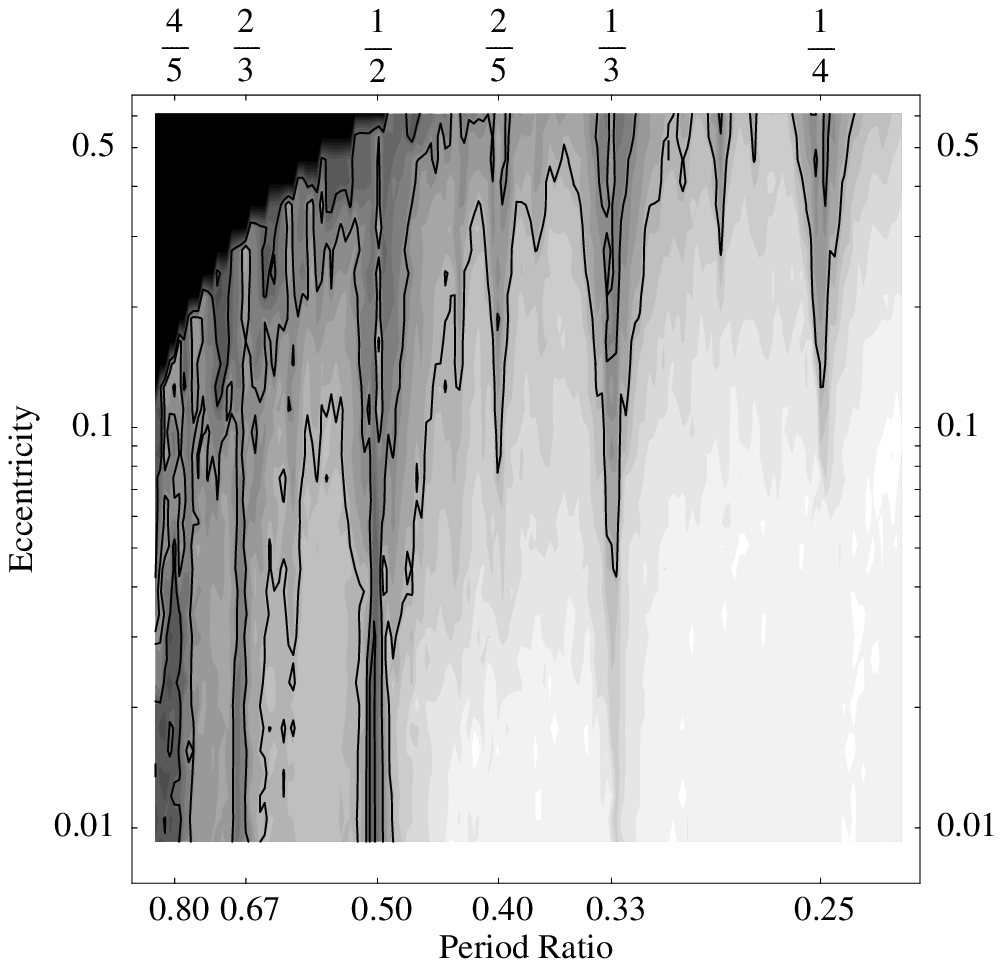}
\caption{The upper limit on the mass of an interior perturbing planet in the TrES-1 system from the 12-point analysis.  The regions bounded by the contours correspond to a companion mass that is less than 100, 10, and 1 earth masses.}
\label{3siginx}
\end{figure}

\begin{figure}
\includegraphics[width=\fsize]{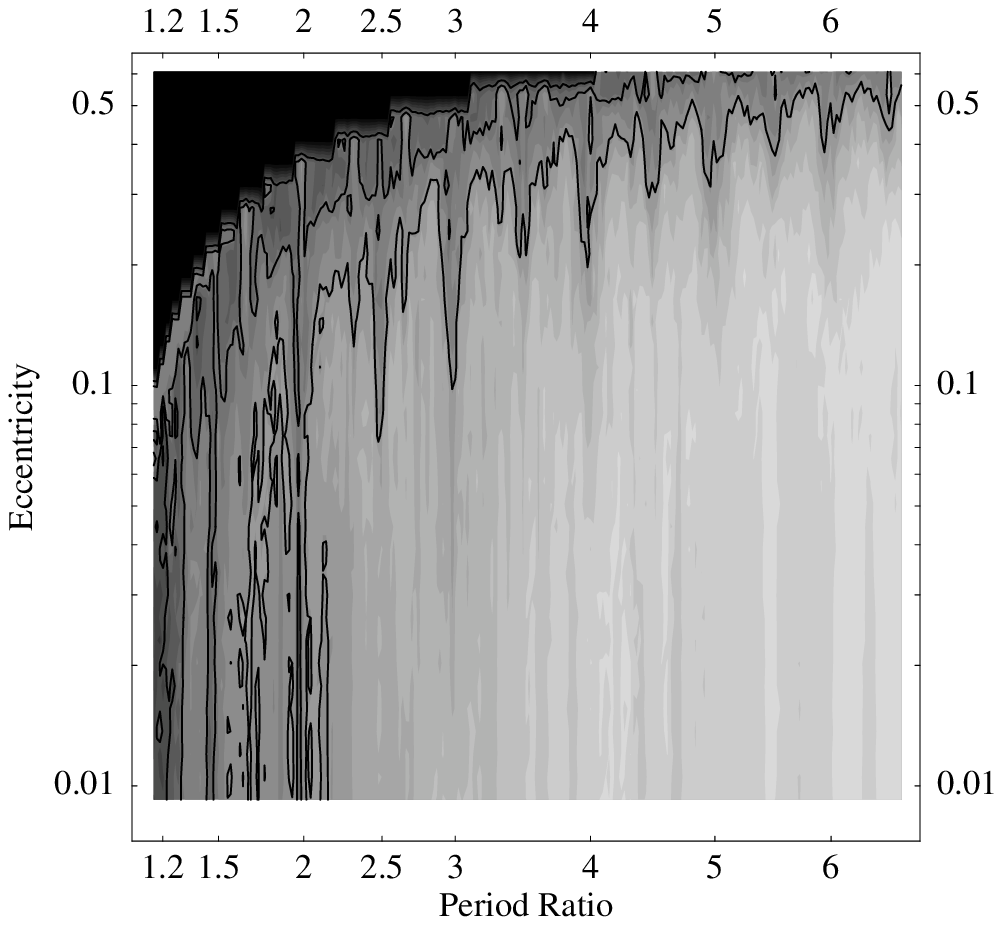}
\caption{The upper limit on the mass of an exterior perturbing planet in the TrES-1 system from the 12-point analysis.  The regions bounded by the contours correspond to a companion mass that is less than 100, 10, and 1 earth masses.}
\label{3sigoutx}
\end{figure}

The mass that corresponds to a 3-$\sigma$ increase in the \cs\ constitutes our estimate for the maximum allowed mass of a secondary planet.  I use the approach outlined by \cite{press} where I locate the mass that causes the difference between the \cs\ of the null hypothesis and the \cs\ obtained with a secondary planet to equal $\Delta \chi^2 = 9$.  This gives the 3-$\sigma$ limit because only one parameter is allowed to vary---the remaining parameters are either fixed or marginalized at each point.  This maximum mass is shown as a function of $a$ and $e$ for the 11-point set in Figures \ref{3sigin} \ref{3sigout} for an interior and an exterior perturber respectively.  The contours correspond to 100, 10, and 1 $M_{\oplus}$.  The dark portion in the upper-left corners are where the orbits overlap and I assign a mass of $10^{-8} M_{\odot}$ to those locations.  We see from these figures many regions where the mass of a companion must be comparable to or less than the mass of the Earth regardless of its orbital eccentricity.  The most stringent constraints are near the $j$:$j+1$ mean-motion resonances.  Of particular interest are the very tight constraints for low-eccentricity perturbers (which constraints continue to an eccentricity of zero) because tidal circularization drives the eccentricity down.  The 12-point analysis gave similar results which shown in Figures \ref{3siginx} and \ref{3sigoutx}.

\section{Comparison with Radial Velocity}

\begin{figure}
\includegraphics[width=\fsize]{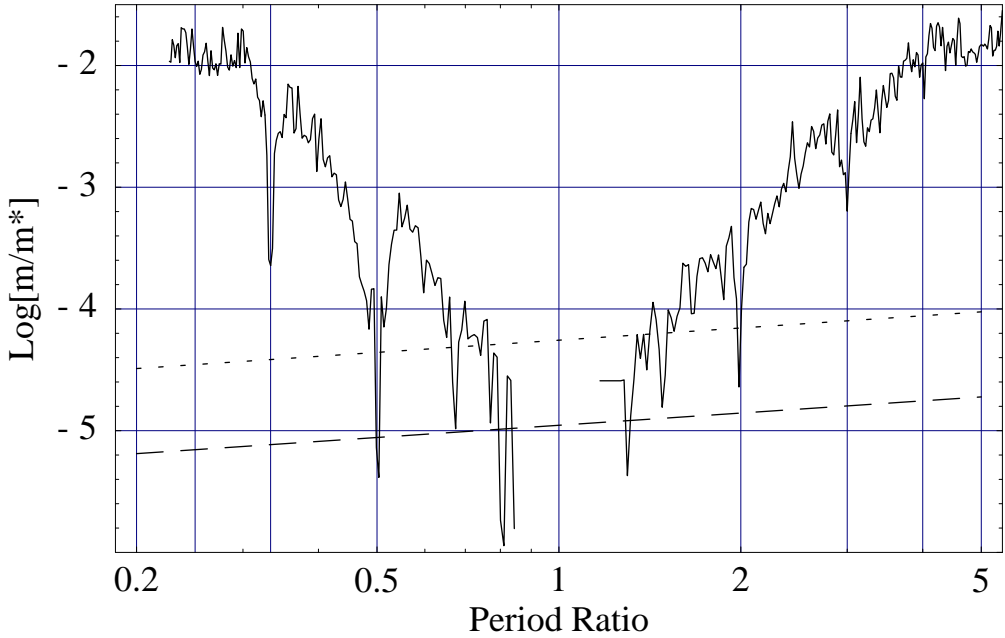}
\caption{A comparison of the limits expressed in this paper for the 11-point analysis (solid curve) and the expected limits that are achieved with 11 randomly selected radial velocity measurements that have a precision of 5 m ~s$^{-1}$ (dotted curve) and 1 m ~s$^{-1}$ (dashed curve) for a secondary planet with an eccentricity of 0.05.  The radial velocity curves are given by equation (\ref{rvform}) using $\Delta \chi^2 = 9$, $N = 11$, $Q_0 = 0$, and $Q = 2$.  Near the 4:3, 3:4, and the 1:2 mean-motion resonances the TTV analysis of these data can place tighter constraints (about a factor of 3) on the mass of putative secondary planets than can the RV technique with 1 m ~s$^{-1}$ precision.}
\label{rvcomp}
\end{figure}

\begin{figure}
\includegraphics[width=\fsize]{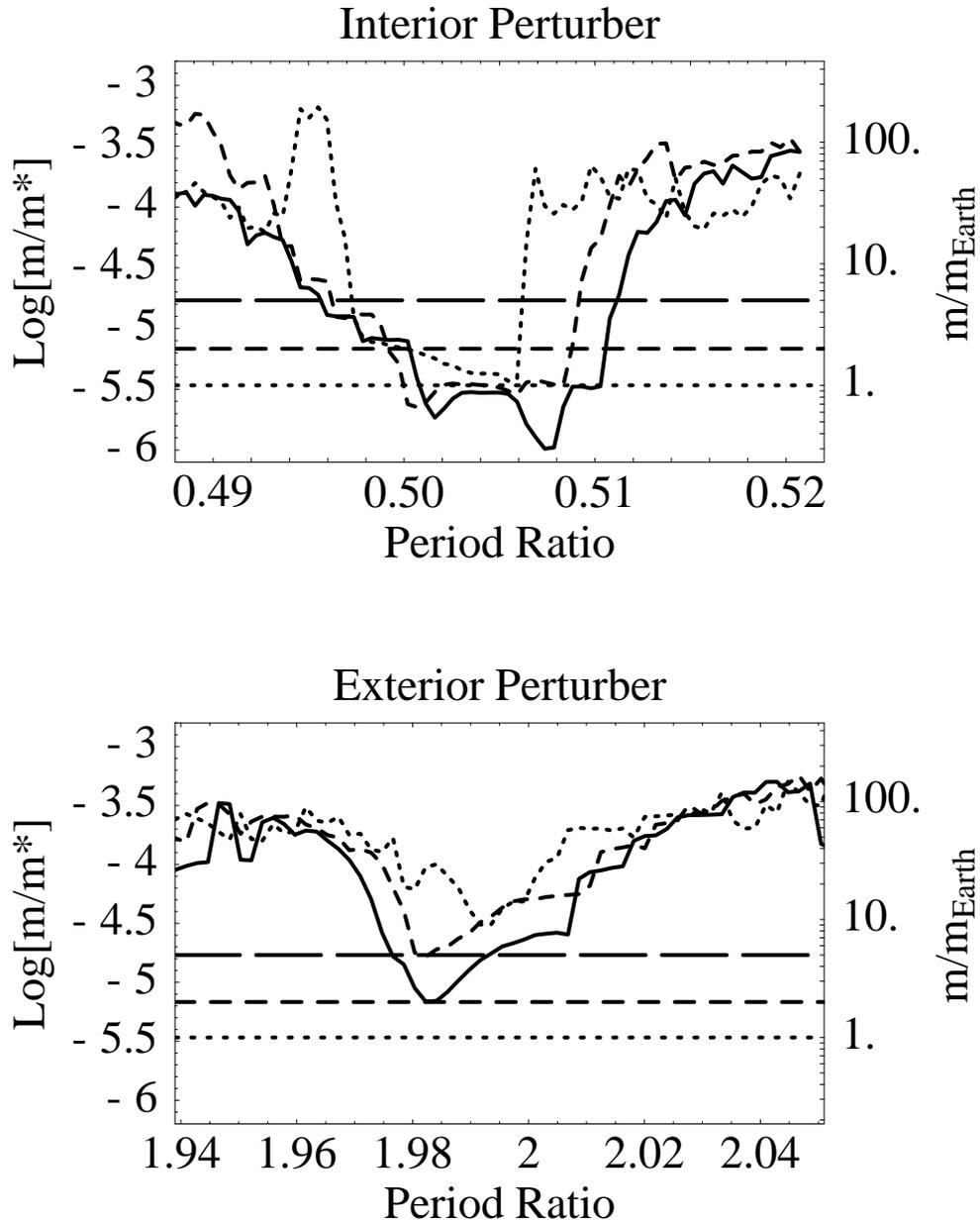}
\caption{This shows the perturber mass limit from a more refined TTV analysis near the 2:1 mean-motion resonances as a function of the period ratio of the perturber and TrES-1b.  The upper panel is for an interior perturber and the lower panel is for an exterior perturber.  The solid curve is for a perturber eccentricity of 0.00, the dashed curve is for an eccentricity of 0.02, and the dotted curve is for an eccentricity of 0.05.  The long-dashed line shows a mass of $5M_{\oplus}$, the dashed line is for $2M_{\oplus}$, and the dotted line is for $1M_{\oplus}$.  For a perturber with eccentricity 0.02, the lowest point on the curve corresponds to a mass of $0.65M_{\oplus}$.}
\label{zoom1}
\end{figure}

\begin{figure}
\includegraphics[width=\fsize]{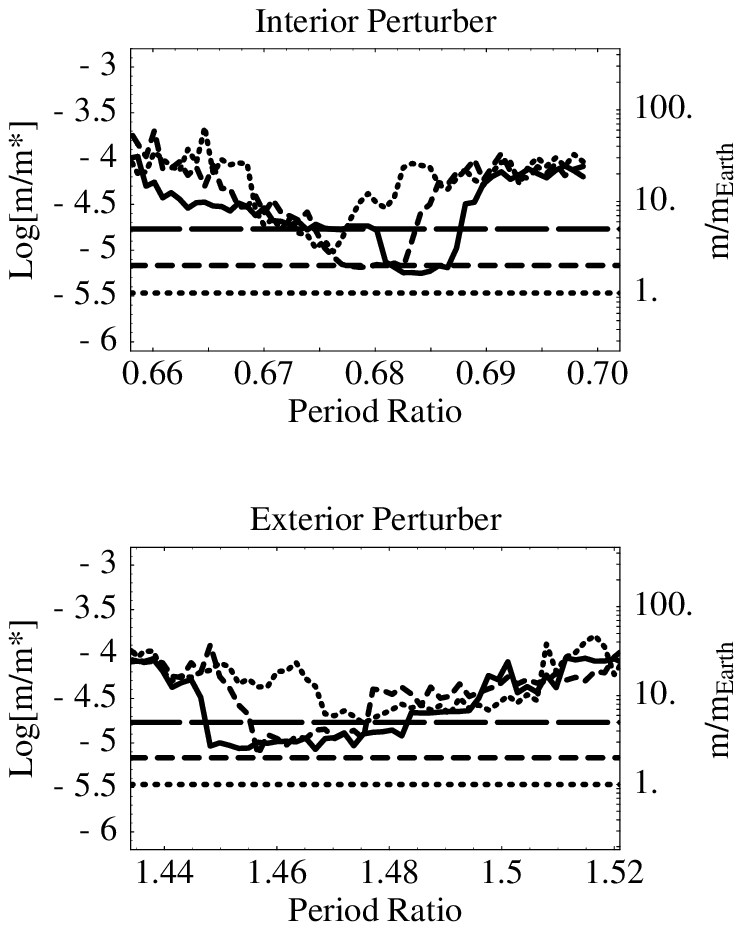}
\caption{This shows the perturber mass limit from a more refined TTV analysis near the 3:2 mean-motion resonances as a function of the period ratio of the perturber and TrES-1b.  The legend for this plot is the same as in Figure \ref{zoom1}.}
\label{zoom2}
\end{figure}

Much of the region where a secondary planet is not tightly constrained by this analysis could be limited more efficiently by radial velocity measurements.  Figure \ref{rvcomp} shows the constraint achieved from this TTV analysis and the constraint from a hypothetical RV analysis, with the same number of data, as a function of the period ratio of the two planets.  I assume that the RV measurements have a precision of both 5 and 1 m ~s$^{-1}$ and that the orbit of the perturber can be treated as circular.  Figures \ref{zoom1} and \ref{zoom2} are the same comparison focussed on the region surrounding the 2:1 and 3:2 mean-motion resonances respectively.

I derived the RV limit in these figures by assuming that the velocity residuals that remain after removing the effects of TrES-1b randomly sample the phase of the putative second planet and that the noise is uncorrelated with the RV measurements.  If no secondary planet exists, then the residuals should surround zero with a variance equal to $\sigma^2$ where $\sigma$ is the precision of the measurements.  The expected \cs\ is then equal to $\chi^2_0 = N-Q_0$ where $Q_0$ is the number of model parameters.  On the other hand, if a secondary planet exists, the expected \cs\ is
\begin{equation}
\chi^2_{\rm p} = \frac{(N-Q)}{\sigma^2} \left[ \frac{\mu^2}{2} \left( \frac{2\pi G M_0}{P} \right)^{2/3} + \sigma^2 \right]
\end{equation}
where $\mu$ is the planet to star mass ratio, $M_0$ is the mass of the star, $G$ is Newton's constant, and $Q$ is the number of model parameters.  By subtracting $\chi^2_0$ from $\chi^2_{\rm p}$ and solving for the mass ratio we obtain the mass of a secondary planet that is detectable with a given $\Delta \chi^2$ threshold as a function of the period of the planet and the precision of the radial velocity measurements
\begin{equation}\label{rvform}
\mu = \sigma \left[ \frac{2(\Delta \chi^2 + Q - Q_0)}{N-Q} \right]^{1/2} \left( \frac{P}{2\pi G M_0} \right)^{1/3}.
\end{equation}

We see from Figure \ref{rvcomp} that for low-order, mean-motion resonances the TTV analysis is able to place constraints on the mass that are nearly an order-of-magnitude smaller than the RV technique with 5 m ~s$^{-1}$ precision and that there are regions where it is more sensitive than RV measurements with 1 m ~s$^{-1}$ precision.  Meanwhile, in nonresonant regimes the RV method remains superior for a large portion of the parameter space.  Additional transit timing data, particularly with precision that is comparable to the most precise of the given data, would lower the entire limit from TTV, provided that no secondary planet exists.  Such data would render the TTV approach superior over a larger range of periods.

\section{Discussion of TrES-1 Results}

Ultimately a combined analysis of all available data, including studies of the stability of candidate systems, will provide the most robust and sensitive determination of the parameters of each planetary system.  None of the planetary systems that compose the limits in Figures \ref{zoom1} and \ref{zoom2} below $10M_{\oplus}$ are stable for more than $10^6$ orbits; though stable orbits with comparable masses, periods, and eccentricities exist.  An overall stability analysis to accompany these TTV analyses was prohibitively expensive.  I estimate that for Figure \ref{3sigin} there were $\sim 10^7$ potential systems that were analysed and $\sim 10^5$ systems were analysed for each curve in Figures \ref{zoom1} and \ref{zoom2}---requiring a total of $\sim 10^8$ hours of processor time.  General stability analyses are often better suited to confirm or study a particular candidate system (e.g. \cite{rory,ji05}) than to constrain candidate systems in the manner that we have done in this section.  We did conduct a stability analysis for the system shown in Figure \ref{timings1} and found it to be unstable within $10^6$ orbits; no neighboring stable systems of comparable \cs\ were found.

It is unclear what fraction of \textit{probable} planetary companions are excluded by our results.  Aside from the fact that many of the known, multi-planetary systems are in mean-motion orbital resonance (e.g. GJ 876), recent works by \cite{thommes}, \cite{papal}, and \cite{zhou} show that, under fairly general initial conditions, small planets are readily trapped in low-order, mean-motion resonances with a gas-giant as it interacts with a protoplanetary disk and migrates inward.  These results may imply that resonant systems are common among multiple-planet systems.  If this is true, then results like Figure \ref{3sigin} actually exclude a much larger fraction of the probable orbits than one might infer from the projection of the excluded regions onto the $a$/$e$ plane.  In addition, \cite{mande} show that a large fraction of existing, terrestrial planets can survive the migration of a Jupiter-mass planet, though only a fraction of the survivors will be in resonance.

The sensitivity of TTV to the mass of a perturbing planet renders it ideal for discovering and constraining the presence of additional planets in transiting systems like TrES-1.  These studies can help determine the ubiquity of multiple planet systems and resonant systems---including the distribution of mass in those systems.  Moreover, TTV analyses of several systems can play a role in identifying the importance of various planet-formation mechanisms.  For example, the presence of close-in terrestrial planets favors a sequential-accretion model of planet formation over a gravitational instability model~\citep{zhou}.

For the case of TrES-1, while a companion planet with a mass larger than the Earth is ruled out in a portion of the available parameter space, there remain large regions where additional planets could reside; this includes mutually inclined orbits which were not studied in this work (I believe that these results are valid provided that the angle of mutual inclination $i$ satisfies $\cos i \simeq 1$).  The 11 timings analysed in this work, excepting the questionable ``12$^{\rm th}$'' point, deviate from a constant period by an amount that is difficult to interpret; having a \cs\ of about 2 per degree of freedom.  Additional observations, with higher precision, will allow us to more fully constrain the system and to interpret the existing transit timing variations.

\chapter{The HD 209458 System\label{hd209}}

We now turn our attention to the analysis of transit data from the HD209458 system.  The observations for HD209458 were reported by \cite{bro01,schult} however, we re-reduced the raw data to obtain the transit times that are analyzed here.  Many of the results of this chapter will be reported in the forthcoming \cite{agol06}.  Since the data from this system were obtained using the Hubble Space Telescope, the precision of the timing measurements is significantly improved over those from the TrES-1 system.  This makes these data capable of probing for yet smaller companions---well below the mass of the Earth in some regimes.  Moreover, we will see that additional steps can be taken during observations to further improve the precision of the timing measurements.

\section{Data Reduction}

At least ten science observing programs with the Hubble Space Telescope
have proposed to observe the transit of the planet in system HD209458.
The five ultraviolet observations do not have enough photons to measure 
the transit lightcurve precisely, while five programs have made 
observations of the planetary transit in the optical and infrared, of 
which three are now publicly
available: 8789, 9171 and 9447 (the other two will become available 
in the next year).

Two programs, Brown 8789 and Charbonneau
9447, utilized the STIS spectrograph as a CCD photometer, spreading
the light from the star using a grism to capture as many photons
as possible.  This approach provides what may be the most precise 
relative photometry
ever \citep{bro01}. 
The third program
utilized the Fine Guidance Sensors (FGS) to carry out rapid high-precision 
photometry \citep{schult}. 
These programs give
a total of thirteen transits observed with HST that are now
available.  A summary of these observations is given in table \ref{tab01}.  
The remaining programs, with data that are not public, will provide another nine transits for future studies.

\begin{table}
\caption{Transit epoch and number, instrument, wavelength range, cadence, and dates for HD209458 transit observations.}\label{tab01}
\begin{center}
\begin{tabular}{@{}lccccll}
Transit & 
    Orbs. & 
       Inst. & 
              Filt. & 
                      wavelengths & 
                                        exp./read & 
                                                    date (UT) \cr
\hline
1 /  0& 5& STIS & G750M & 581.3-638.2 nm &  60/20 sec & 2000 Apr 25 \cr
2 /  1& 5& STIS & G750M & 581.3-638.2 nm &  60/20 sec & 2000 Apr 28-29 \cr
3 /  3& 5& STIS & G750M & 581.3-638.2 nm &  60/20 sec & 2000 May 5-6 \cr
4 /  5& 5& STIS & G750M & 581.3-638.2 nm &  60/20 sec & 2000 May 12-13 \cr
5 /117& 3& FGS  & F550W & 510-587.5 nm   &0.025/0 sec & 2001 Jun 11 \cr
6 /143& 2& FGS  & F550W & 510-587.5 nm   &0.025/0 sec & 2001 Sep 11 \cr
7 /160& 2& FGS  & F550W & 510-587.5 nm   &0.025/0 sec & 2001 Nov 10 \cr
8 /179& 2& FGS  & F550W & 510-587.5 nm   &0.025/0 sec & 2002 Jan 16 \cr
9 /252& 2& FGS  & F550W & 510-587.5 nm   &0.025/0 sec & 2002 Sep 30 \cr
10/313& 5& STIS & G430L & 290-570   nm   &  22/20 sec & 2003 May 3 \cr
11/321& 5& STIS & G750L & 524-1027  nm   &  19/20 sec & 2003 May 31 \cr
12/328& 5& STIS & G430L & 290-570   nm   &  22/20 sec & 2003 Jun 25 \cr
13/331& 5& STIS & G750L & 524-1027  nm   &  19/20 sec & 2003 Jul 5-6 \cr
\end{tabular}
\end{center}
\end{table}

The primary science objectives for these observations 
were to constrain the planetary parameters,
to detect absorption features in the planet's atmosphere, and
to search for satellites of the planet.  Besides these studies, the data are
useful for timing each transit to measure a precise ephemeris \citep{wit05}
and for a TTV analysis.  However, in order to obtain the transit times from each set of observations we needed to re-reduce the pipeline calibrated photometric data that is available in the archive.  This process involved several steps; 
following in part 
the reduction procedures described in \cite{bro01} 
and \cite{schult}. 
I summarize the most important steps here for each data set.
For a more detailed description of the data reduction procedure see \cite{agol06} where the reduction is described in some detail.

\subsection{Outlier Removal and Data Binning}

For STIS we carried out the minimum reduction necessary to obtain
precise photometry.  We subtracted cosmic-rays using
a 5 sigma rejection of the time series for each pixel within an
HST orbit, we binned the photons within 8 pixels of the spectrum 
peak (17 pixels wide) to derive the total counts for each frame, 
and assigned the midpoint of each exposure as the time for each frame. 
We discarded the first frame from each orbit and for the first 
transit observed in program 8789 we only used the data for which the 
spectrum centroid was greater than 4 pixels from the edge of the CCD 
\citep[see discussion in][]{bro01}; 
in addition we did not
utilize the first and last columns of the CCD which show larger
errors.  The 1-sigma errors on the flux are taken from the pipeline
calibrated errors which are summed in quadrature.

For FGS we took the sum of the four photometers, discarded
data which deviated more than 3 sigma from a fourth-order polynomial fit
to the time series for each orbit and binned the
data to 80 second bins (10 second bins gave the same results)
with the time stamp at the center of each bin.  
With these time series we then fit a model for the transiting
planet and the flux sensitivity variations of Hubble, which I
describe later.

\subsection{Photometric Error bars}

The errors computed from the pipeline include counting noise and,
for the CCD data, read noise.  We compared the scatter in the data 
to the pipeline error bars and found that the scatter was
larger than the pipeline errors for the FGS and second set of STIS data.
This could not be attributed solely to cosmic rays and we were unable 
to identify the additional source of noise.  
Therefore, we estimate the photometric errors 
for each individual transit by analyzing the out-of-transit data.

To account for the fluctuations in sensitivity of HST when
determining the size of the error bars, we fit a model to the 
out-of-transit data for each transit with two components:  i) a 
5th order polynomial fit as a function of the orbital phase of HST 
and ii) a linear fit as a function of time.
The residuals did not correlate with
the HST orbital phase so we assigned the greater of the standard
deviation of the residuals for the polynomial fit or the statistical error.  
The first five transits had statistical errors nearly equal to the residuals,
while the error bars of the remaining transits were increased by 20-50\%.
These changes to the photometric error bars were used throughout the remainder
of the analysis.



\subsection{Limb darkening}

If the lightcurves for each transit were only affected by
Poisson noise and were continuously sampled, then assumptions
about the shape of the transit lightcurve would be unnecessary:  
the mid-point of transit could be found by assuming the
transit lightcurve was symmetric.  However, because the Earth occults 
the Hubble Space Telescope, the data are
affected by sensitivity variations due to temperature changes
of the telescope and by gaps in the observations.  
We must, therefore, choose a model for the 
transit lightcurve in order to identify the midpoint of the transit---the 
``transit time''.  The primary uncertainty in the modeling is the
shape of the stellar limb darkening.  We modeled the limb-darkening 
of the star with two approaches: i) stellar atmosphere predictions and 
ii) quadratic limb-darkening.  By comparing the results from these 
two approaches we test the robustness
of the transit times under changes to our assumptions about limb-darkening,

Despite the differences
between these these two approaches, the measured times were
nearly identical within the errors and yielded a similar 
reduced $\chi^2$.  However, the inferred planetary radius
and orbital inclination differed by more than the errors
for each model. 

\subsection{HST flux sensitivity variations}

The thermal fluctuations of HST during its orbit lead to changes in
the sensitivity of each instrument at the 0.5\% level.  Fortunately
these variations appear to be 
reproducible during
successive orbits (in addition to a smaller linear drift
between successive orbits), but they are not reproducible on longer 
timescales.  Following \cite{bro01} 
and \cite{schult} 
we phased the observations of each transit to the HST orbital period 
($p_h$).  We used two models for the variation of the flux
sensitivity with orbital phase and two models for the ``secular" variation
with time over each transit, combined together giving a total
of four models for the background variations.

We modeled the sensitivity variations as a function of the orbital phase 
of HST as either a fifth-order polynomial, $\sum_{i=0}^4 c_i 
\phi^{(i+1)}$, where $0\le\phi<2\pi$ is the orbital phase or
as a linear and two harmonic functions 
$c_0\phi+c_1\sin(\phi)+c_2\cos(\phi) +c_3\sin(2\phi)+c_4\cos(2\phi)$.
The secular variation with time we modeled either as a linear function,
$c_5 f_{tran}(t)+c_6 t$, where $f_{tran}$ is the transit lightcurve
model normalized to unity outside transit or as a constant coefficient 
for each transit, $c_{i+5} f_{tran,i}(t)$ where $i=0,12$ labels each 
transit.

The resulting lightcurve model depends on sixteen non-linear parameters (the 3 
planetary orbital parameters and 13 transit times).  For the
models of the HST flux sensitivity variations there are either 91 or 
113 linear parameters (6-9 for the HST sensitivity for each transit).
Coincidentally, the physically interesting parameters are non-linear,
while the uninteresting parameters are linear.  
We mapped the
$\chi^2$ as a function of each non-linear parameter (while marginalizing
over the other parameters) and found that the $\chi^2$ has a quadratic
shape near the minimum---indicating a unique solution.





\subsection{Timing Error Computation}


The best-fit transit timings for the STIS data have reduced 
chi-square in the range 
$\bar\chi^2=1.07-1.25$ for the four different models of the flux sensitivity.  
However, the reduced chi-squares
for the FGS data (which contain fewer data points) were as high as 2.6
(for the 5th transit) indicating that the models were a poor fit to
the data.  We examined the residuals of the FGS data and found that some 
residuals within the transits were correlated.  
Thus, our best fitting models were not a proper description of
the data---presumably due to either errors in the limb darkening 
or in the modeling of the HST flux sensitivity variations.  
In effort to incorporate these systematic effects into our estimate 
of the photometric error bars we used a Monte
Carlo bootstrap simulation of the errors where, in order to maintain 
the correlations, we shifted the residuals for each transit by a 
random number of points.
We also reversed the residuals for each orbit and repeated the shift.  This
process was repeated 200 times; each time adding 
the shifted residuals to the best-fit
model and then re-fitting with the same procedure that was applied to the
original data as described in the previous section.  Our expectation is that 
this procedure will give a better estimate of the errors that are 
due to inaccuracies in our assumed flux sensitivity models.  
We found that the resulting errors 
were larger by a factor of a few than if we simply used Poisson errors for the 
bootstrap or resampled the residuals randomly.

Although this process gives an indication of the size of the systematic errors,
we also assessed the systematic errors by comparing the results from the
four different HST flux sensitivity variation models.  The differences 
in $\chi^2$ between the four different models was small---all 
had reduced $\chi^2$ near unity, so it is difficult to favor one
particular sensitivity variation model.  However, the differences between
the derived transit times for each model was larger than the size of the 
error bars for a given model.  We believe that this discrepancy is
probably due to imperfections in all of the four flux sensitivity models,
so we have taken the mean and standard deviations of all eight hundred
Monte-Carlo simulations as an estimate of the transit times and their
errors.  Finally, to check the robustness of the stellar atmosphere
limb-darkening we have used the quadratic limb-darkening law with
the polynomial function of phase and linear secular term to fit
the transit times.  We find that these times agree within the errors 
with the times derived using the stellar atmosphere model and the same flux 
sensitivity model, so we conclude that the limb-darkening model
provides an accurate fit to the data.

To these measured transit times we apply a correction for the
motion of the Earth about the barycenter of the solar system
(a heliocentric correction is insufficient as it differs from
barycentric by up to 5 seconds).  The motion of the Hubble Space Telescope 
about the Earth contributes a negligible timing error.
The resulting transit times are reported in Table \ref{tab02}.
We find from the Monte Carlo simulations that there is no significant
correlation between the individual transit times so we report the 
standard deviation of each time without the full covariance matrix.

\begin{table}
\caption{Transit epoch, transit time, and timing uncertainty for HD 209458 transit observations.}\label{tab02}
\begin{center}
\begin{tabular}{@{}lccc}
Transit & Transit time    & Error & Error     \cr
        & (- 2450000 BJD) & (days)&(seconds) \cr
  1 &    1659.93678 &    0.00015 &  12 \cr
  2 &    1663.46150 &    0.00025 &  21 \cr
  3 &    1670.51102 &    0.00013 &  11 \cr
  4 &    1677.56044 &    0.00037 &  31 \cr
  5 &    2072.33284 &    0.00049 &  42 \cr
  6 &    2163.97574 &    0.00028 &  24 \cr
  7 &    2223.89685 &    0.00058 &  50 \cr
  8 &    2290.86741 &    0.00051 &  44 \cr
  9 &    2548.17311 &    0.00029 &  24 \cr
 10 &    2763.18306 &    0.00018 &  15 \cr
 11 &    2791.38087 &    0.00013 &  10 \cr
 12 &    2816.05431 &    0.00021 &  18 \cr
 13 &    2826.62867 &    0.00017 &  14 \cr
\end{tabular}
\end{center}
\end{table}

\begin{figure}
\includegraphics[width=\fsize]{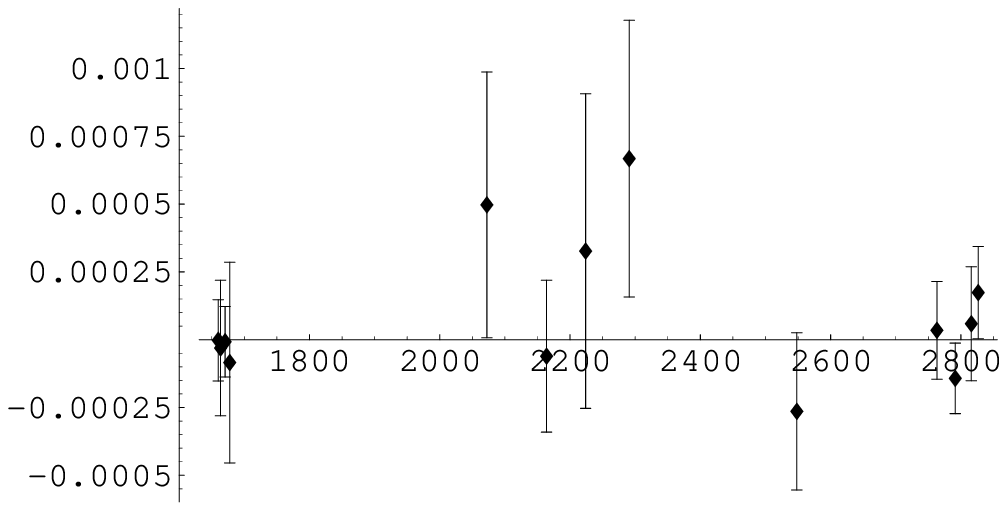}
\caption{This shows the deviations from a constant period of the transit times for the HD209458 system.  The horizontal axis is in $HJD - 2450000$ and the vertical axis is in days.}
\end{figure}

\section{Ephemeris}

The accurate and realistic estimation of errors on the transit times
allows for an accurate estimate of the ephemeris of the transiting
planet.  We find values of:  $T_0=2451659.93677 \pm  0.00009$ BJD
(error is 8 seconds)
and $P=3.5247484  \pm 0.0000004$ days  (error is 0.03 seconds).
The period and zero-point of the ephemeris are anticorrelated with a 
correlation coefficient of -0.8;  this is due to the fact that an
increased zero point requires the period to shrink to go through
the later data points.  The anti-correlation decreases to -0.2 if
the zeroth eclipse is chosen to be halfway between the first and last
eclipses.  Given the estimated statistical plus
systematic errors, the chi-square is $\chi^2 = 6.0$ for eleven
degrees of freedem (thirteen transits minus two fitting parameters).
This indicates that the transit times are consistent with being
uniform and indicates that there may not be a second planet
detectable with the current HST data.  However, as with the TrES-1 system
these data can be used to place constraints on the masses of other 
planets in the system.

\section{Constraints on Secondary Planets}

\begin{figure}
\includegraphics[width=\fsize]{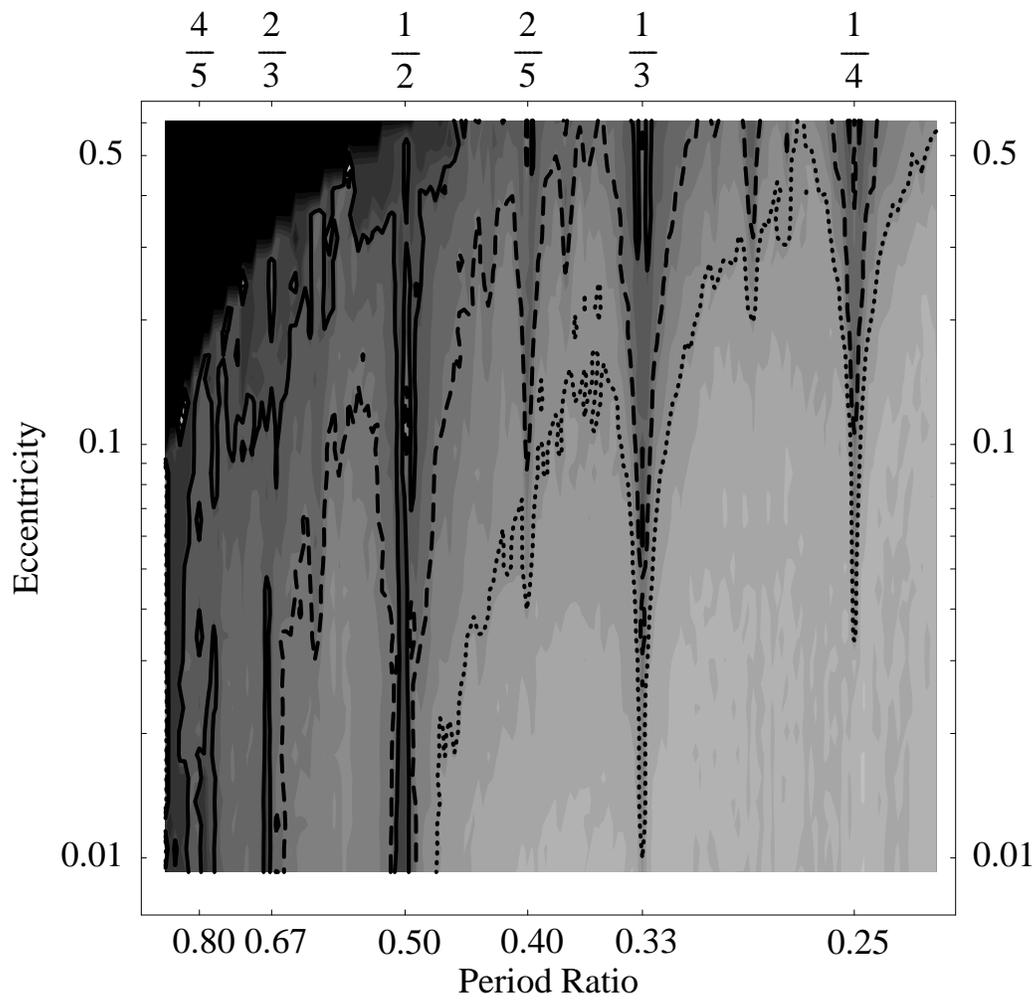}
\caption{The upper limit on the mass of an interior perturbing planet in the HD209458 system.  The regions bounded by the contours correspond to a companion mass that is less than 100 (dotted), 10 (dashed), and 1 (solid) earth masses.}
\label{hdin}
\end{figure}

\begin{figure}
\includegraphics[width=\fsize]{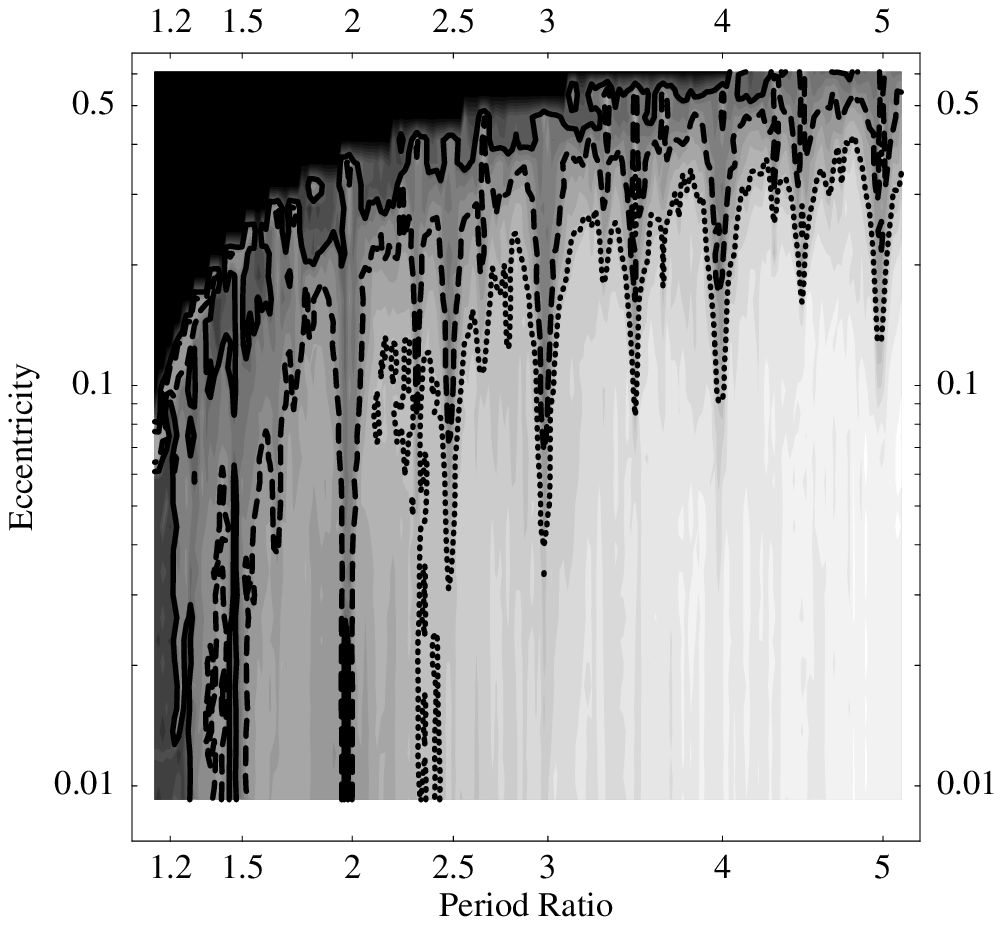}
\caption{The upper limit on the mass of an exterior perturbing planet in the HD209458 system.  The regions bounded by the contours correspond to a companion mass that is less than 100 (dotted), 10 (dashed), and 1 (solid) earth masses.}
\label{hdout}
\end{figure}

I conducted an analysis similar to that for the TrES-1 system for the HD209458 system using the transit times given in table \ref{tab02}.  However, because the reduced \cs\ for these transit times is less than unity, I did not attempt to find the best-fitting two-planet system since any improvement in the overall reduced \cs\ would be suspect.  The constraints on the mass of a secondary planet in the HD209458 system are shown in Figure \ref{hdin} for an interior perturber and in Figure \ref{hdout} for an exterior perturber.

\begin{figure}
\includegraphics[width=\fsize]{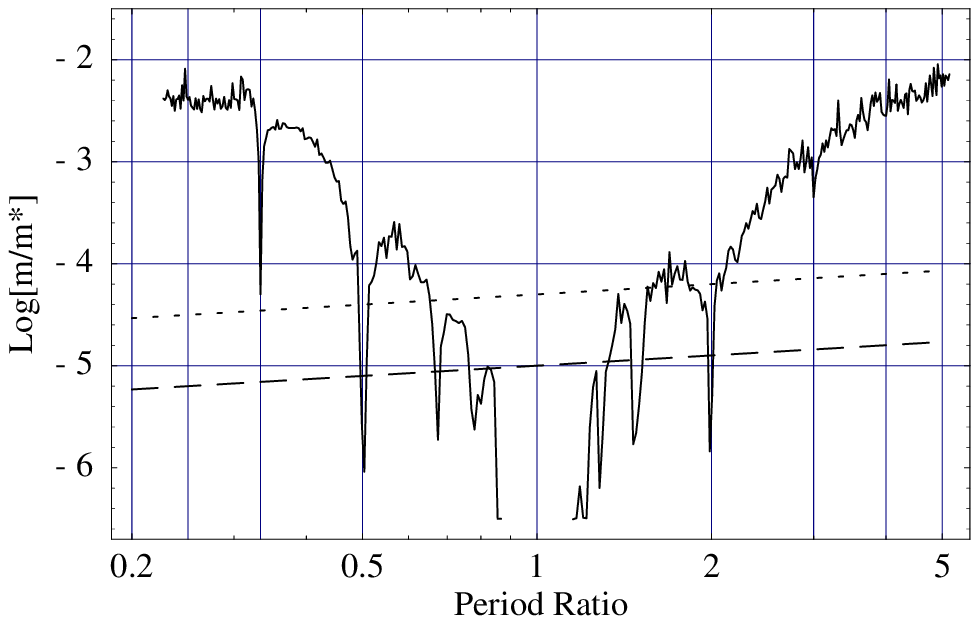}
\caption{A comparison of the limits expressed in this paper (solid curve) and the expected limits that are achieved with 13 randomly selected radial velocity measurements that have a precision of 5 m ~s$^{-1}$ (dotted curve) and 1 m ~s$^{-1}$ (dashed curve) for a secondary planet with an eccentricity of 0.05.  The radial velocity curves are given by equation (\ref{rvform}) using $\Delta \chi^2 = 9$, $N = 13$, $Q_0 = 0$, and $Q = 2$.  Near the 4:3, 3:4, and the 1:2 mean-motion resonances the TTV analysis of these data can place tighter constraints (about a factor of 3) on the mass of putative secondary planets than can the RV technique with 1 m ~s$^{-1}$ precision.}
\label{comphd}
\end{figure}

As expected (in light of the results for the TrES-1 system) near the low-order mean-motion resonances the maximum allowed mass is comparable to or less than the mass of the Earth.  A comparison to the radial velocity technique, similar to Figure \ref{rvcomp}, is shown in Figure \ref{comphd} where we see that both the increase in the number of data (13 instead of 11) and the increased timing precision (and average of 25s instead of 108s) renders the TTV technique superior to the radial velocity technique over a larger range of period ratios.

\section{Statements About Space and Ground Based Observations}

\begin{figure}
\includegraphics[width=\fsize]{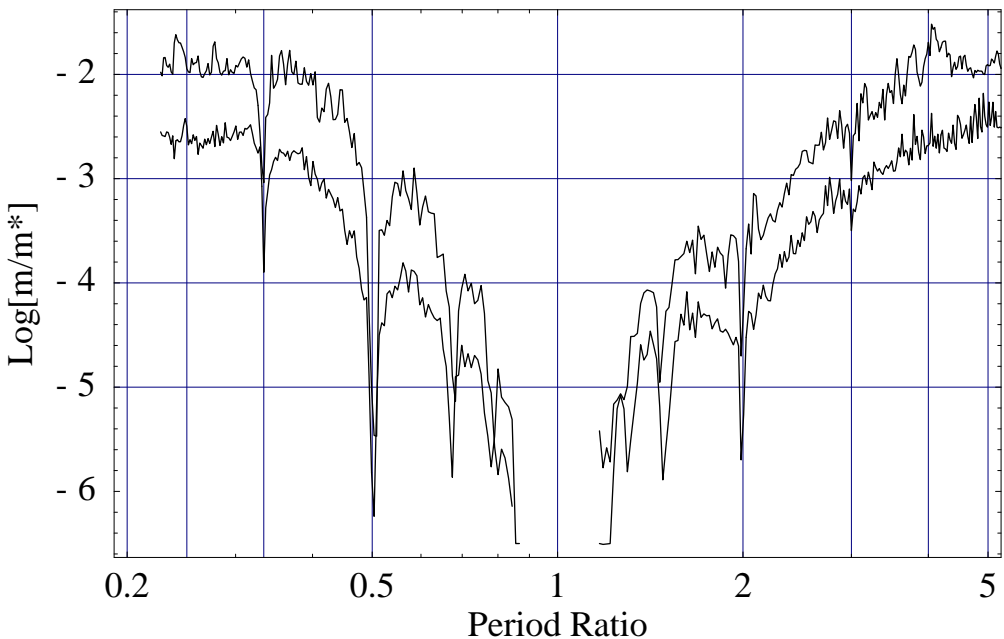}
\caption{This image shows the mass limit of a secondary planet obtained from a TTV analysis of the 11 transits for the TrES-1 system (upper curve) with the limit obtained from the 13 transits of the HD209458 system (lower curve).}
\label{comphdt1}
\end{figure}

A comparison of the results for the TrES-1 system and for the HD209458 system demonstrate the power that the TTV detection technique has to constrain companions in transiting systems using either ground-based observations or space-based observations, or both.  It is somewhat surprising that ground-based observations taken with modest telescopes are able to probe for such small planetary companions.  The average timing uncertainty in the data from the TrES-1 system was 108s compared to 25s for the HD209458 system.  This factor of four in precision resulted in a comparable tightening of the constraints on the presence of secondary planets in the HD209458 system (see Figure \ref{comphdt1}).  However, the most precise of the data for the TrES-1 system, taken with the 1.2m telescope at the Whipple Observatory, had an uncertainty of 26s---almost identical to the average of the space-based observations.  Moreover, the ground-based observations were not optimized for determining the time of the transit.  This shows that ground based observations can play an important role in the collection of transit data for TTV analysis.  I note that the STIS observations, with an average uncertainty of 17s, consistently gave better timing precision than the FGS data and is therefore of higher quality than the best ground-based observations of TrES-1.  However, the potential for high quality observations from the ground to probe for terrestrial planets remains.

While space-based observations will almost always provide higher precision transit times than ground-based observations, the availability of 1-2m telescopes is significantly higher than the availability of the few space-based observatories.  Thus, ground-based transit observations will likely provide follow-up observations of planetary systems that are discovered by the \textit{CoRoT} and \textit{Kepler} missions; and entirely ground-based transit programs such as \textit{XO} certainly have the capacity to provide data of sufficent quality for interesting TTV analyses.

Another item that became clear during the reduction of the HD209458 transit data is that despite its lofty position above the Earth's atmosphere, it is 
apparent that the Hubble Space Telescope is not
ideally suited to timing transits.  The main problem is that its
low-Earth orbit causes occultation of the Sun which leads to
thermal variations within the orbit and causes occultation of
the target star which leads to gaps and repointing.  The thermal
variations cause sensitivity variations, while the gaps make
a model for limb-darkening necessary to measure the transit
times.  If the data errors behaved according the Poisson statistics,
the expected time variations for the STIS data would be of order 
3-5 seconds, while the derived times are 2-3 times larger.  The
FGS data had a decreased sensitivity relative to the STIS data
due to a lower count rate and the observation of each transit
for only two or three orbits, an observing strategy which conserved
on Hubble time but was unsuited to the precise measurement of transit 
times.  Future precise measurements of transit times are possibly 
better carried out with satellites in Earth-trailing orbit, such
as the Spitzer Space Telescope.  Spitzer has the extra advantage
that the infrared is less subject to limb-darkening; however, it
will collect fewer photons because of its smaller aperture and because it 
observes past the peak of the photon count spectrum of G-type stars.

\chapter{Conclusion\label{conclude}}
The results presented in this dissertation favor the thesis that by analyzing the variations in transit times of a transiting extrasolar planet one can extract important scientific information.  This information may include the orbital elements of unseen, perturbing planets---or, constraints on the presence of such planets, the absolute mass and radii of the bodies in the system, and relevant facts that bear on the formation and evolution of planetary systems.  Thus far the first TTV analysis, that of the transits of the TrES-1 system, constitutes the first, published probe for sub earth-mass planets orbiting a distant, main-sequence star.  While no discovery was made, the constraints that it places on the presence of secondary planets in the important, low-order orbital resonances remains an important result because if the implications that it has for theories of planet formation~\cite{zhou}.  The same can be said for the similar analysis---with similar results---of the transits from the HD209458 system.

However, two systems with similar results does not a trend make.  Before I can make definitive statements concerning the results from TTV analyses I need to analyze many more systems; systems that should result from existing and planned transit searches (e.g. \cite{mcc05,borucki,bourde03}).  At this time, much development remains before TTV can be systematically applied to the large amounts of transit data that are expected over the next decade.

The research that remains has observational, theoretical, and analysis aspects.  Some of these topics include: 1) optimizing ground and space-based observations for the timing of planetary transits and characterizing the effects of coverage gaps and timing noise on the observational side; 2) theoretical development to understand the nature of $j$:1 mean-motion resonances, the effects of mutually inclined orbits, the long-term stability of multiple planet systems, and the consequences of additional planetary companions beyond two; and 3) optimizing the existing analysis software so that it runs more efficiently and that it is can be applied more readily to multiple systems instead of the current situation of one system at a time.

As new data from transiting systems becomes available TTV should prove an important tool in addressing some of the fundamental questions concerning planetary systems in general and in making important discoveries in multiple planet systems.  This is in part because of the sensitivity of the TTV scheme to the mass of a perturbing planet---it being capable of probing for terrestrial planets with data taken from modest, ground-based telescopes.  It is also because of TTV's unique ability to break existing degeneracies in the parameters involved in planetary systems such as the mass/radius degeneracy mentioned earlier.  Also, TTV is particularly well-suited to probe for small objects in mean-motion resonances and to possibly determine the relative inclinations of planetary systems.

%
%
\setlength{\baselineskip}{0.6\baselineskip}
\bibliographystyle{plain}
\bibliography{dissertation}
\setlength{\baselineskip}{1.667\baselineskip}

%
%
\appendix
\raggedbottom\sloppy
 
 
 
\vita{
Jason Hyrum Steffen was born in Fairfield, California on May 15, 1975 to David and Thail Steffen.  He moved to Price, Utah when he was a few months old.  After a brief excursion to Orangeville, Utah he moved to Fruit Heights, Utah where he attended grade through high school---graduating in 1993 with High Honors from Davis High School where he participated in almost everything that related to music and percussion.  He then served a two-year mission for the Church of Jesus Christ of Latter-Day Saints in Brussels, Belgium.  The day prior to his return to the United States, he had a brief, uneventful conversation with one Faith Kimball who, two years later, he met again.  A more lengthy series of eventful conversations ensued and they were married in Salt Lake City, Utah on March 11, 2000.  On May 16, 2005 in Kirkland, Washington they were joined by James Hyrum Steffen, an invaluable addition to the family.

Jason attended Weber State University in Ogden, Utah where he graduated \textit{Summa Cum Laude} in Physics and Mathematics in June of 2000.  For a year he worked as a Pricing Analyst for McLeod USA, an Embedded Software Engineer for L-3 Communications, and as a part-time faculty member of the Mathematics department at Salt Lake Community College.  In 2001 Jason enrolled in the graduate school at the University of Washington where he studied Physics with Julianne Dalcanton, Paul Boynton, and finally Eric Agol.  While there he occasionally taught Physics and Astronomy at Edmonds and Bellevue Community Colleges.  His plan upon completion of the requirements for his Doctor of Philosophy in Physics is to move to the Chicago area where he accepted the Brinson Postdoctoral Fellowship offered by the Particle Astrophysics Center at the Fermi National Accelerator Laboratory.
}

\end{document}